\documentclass{iopjournal}

\usepackage{amsmath}
\usepackage{appendix}
\usepackage{xcolor}
\usepackage{soul}
\usepackage{lineno}

\renewcommand\hl[1]{#1} 
\sethlcolor{green}

\begin{document}

\articletype{Paper} 

\title{
Cohort Organized Learning: Clustering Through Agreement
}

\author{
Finn H. O'Shea$^{1, *}$\orcid{0000-0003-2398-7381},
Maria Elena Monzani$^{1,2}$\orcid{0000-0002-8254-5308}
}

\affil{$^1$SLAC National Accelerator Laboratory, 2575 Sand Hill Road, Menlo Park, CA 94025-7015, USA}
\affil{$^2$Kavli Institute for Particle Astrophysics and Cosmology, Stanford University, Stanford, CA 94305-4085 USA}


\affil{$^*$Author to whom any correspondence should be addressed.}

\email{foshea@slac.stanford.edu}


\begin{abstract}

In this article we describe Cohort Organized Learning (CoOL), a method for clustering data without explicit distance or similarity computations.
Herein, we will describe CoOL, derive the gradients determined by expectation maximization to train the networks, show how to monitor convergence during training and evaluate the clusters after training, and discuss a series of examples and use cases.
We also discuss CoOL's limitations and future prospects on related tasks.
Because CoOL uses neural networks to estimate the clusters, it can be used to cluster any data that can be made compatible and we illustrate this on vector data and images.

\end{abstract}

\section{Introduction}
\label{sec:intro}

The modern scientific process produces enormous amounts of data, and most of that data is unlabeled because there is not enough human-hours to sort through it, to say nothing of the cost.
One of the contributors to this prodigious data production is the high dimensionality (i.e. number) of signals recorded during routine operation.
Examples of high-dimensional tools in physics include tokamaks \cite{mcharg_diii-d_2002, abla_iterdb_2014, brans_petabytes_2020}, dark matter detectors \cite{aalbers_data_2024, aprile_xenon_2024}, particle accelerators \cite{thayer_massive_2024, schwarz_aps_data_2019}, telescopes \cite{ivezic_lsst_2019, hasan_prototyping_2025}, and collider detectors \cite{atlas_performance_2009, bernauer_scientific_2023}.
These tools, and many more, each have a trove of potentially useful information that is too costly to organize using conventional methods.
In this article, we resort to clustering, which can be used to partition unlabeled data or to better understand its structure \cite{James_introduction_2013}.
As a fundamental strategy for organizing data, it is used in practically every field that produces data \cite{jain_data_1999, saxena_review_2017, yin_rapid_2024}.

All clustering algorithms, deep-learning included, rely on some measure of distance (or similarity) between the samples\footnote{See Section \ref{sec:relatedwork} for examples of how distance and similarity are used in deep clustering.}.
Herein, we add another strategy by introducing Cohort Organized Learning (CoOL), a method for partitioning data that does not rely on distance or similarity matrix computations between samples in the objective function, but rather uses agreement.
Several differentiable observers (neural networks in our case) use expectation maximization (EM) \cite{dempster_maximum_1977} to reconcile their cluster labels for the samples and learn from each other to partition the samples as a cohort.


Herein, we will describe CoOL, derive the gradients determined by EM to train the observers, show how to monitor convergence during training without labels, describe how to evaluate the success of clustering, and discuss a series of examples using the MNIST handwritten digits dataset \cite{lecun_mnist_1998}.
We are not interested in images per se or learning representations therefrom, so we do not spend any effort attempting to surpass state of the art performance on these datasets.
Along the way, we will discuss some of CoOL's limitations, as we understand them, and how variations on CoOL might be used to solve other problems.

\section{Description of CoOL}
\label{sec:description}


CoOL functions by having a user-defined number of observers estimate the cluster identity of the samples in collaboration with each other.
The mathematical basis for this collaboration is the expectation maximization (EM) \cite{dempster_maximum_1977} work of Dawid and Skene \cite{dawid_maximum_1979}.
In that work, the authors find estimates for the unknown class identity of samples by reconciling the (fixed) labels created by a number of experts.
In doing so, they find estimates for the proportion of samples that belong to each class and the ``error-rate matrix'' (what we will call the reliability matrix) for each observer.

\begin{figure}
 \centering
        \includegraphics[width=0.5\textwidth]{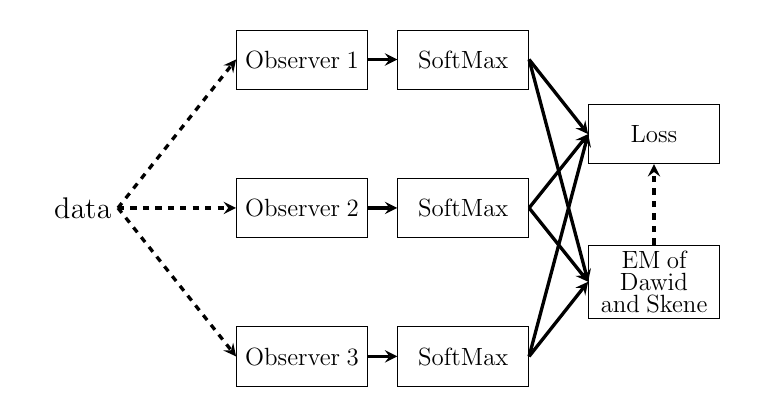}
 \caption{Diagram of data flow through CoOL.  The solid lines show relationships that allow back propagation.}
\label{fig:cool_diagram}
\end{figure}

CoOL modifies this approach, it uses EM and differentiable observers capable of learning to force the observers to agree on the cluster identity of each of the samples.
In other words, the different observers learn to collaboratively cluster the samples as a cohort.
The procedure is straightforward, a diagram is given in Figure \ref{fig:cool_diagram}.
First, the user defines the differentiable observers, the number of clusters those observers should use, and hyperparameters such as learning rate and weight decay \footnote{We do not use weight decay in the work presented here.}. 
Second, the observers are shown the samples (one epoch is showing all the samples to the observers) and each observer produces a label for each sample.
Third, those labels are used to perform one or more steps of the EM-algorithm.
Finally, the results of the EM-algorithm are used to compute a loss for back propagation.


\begin{figure}
 \centering
        \includegraphics[width=0.5\textwidth]{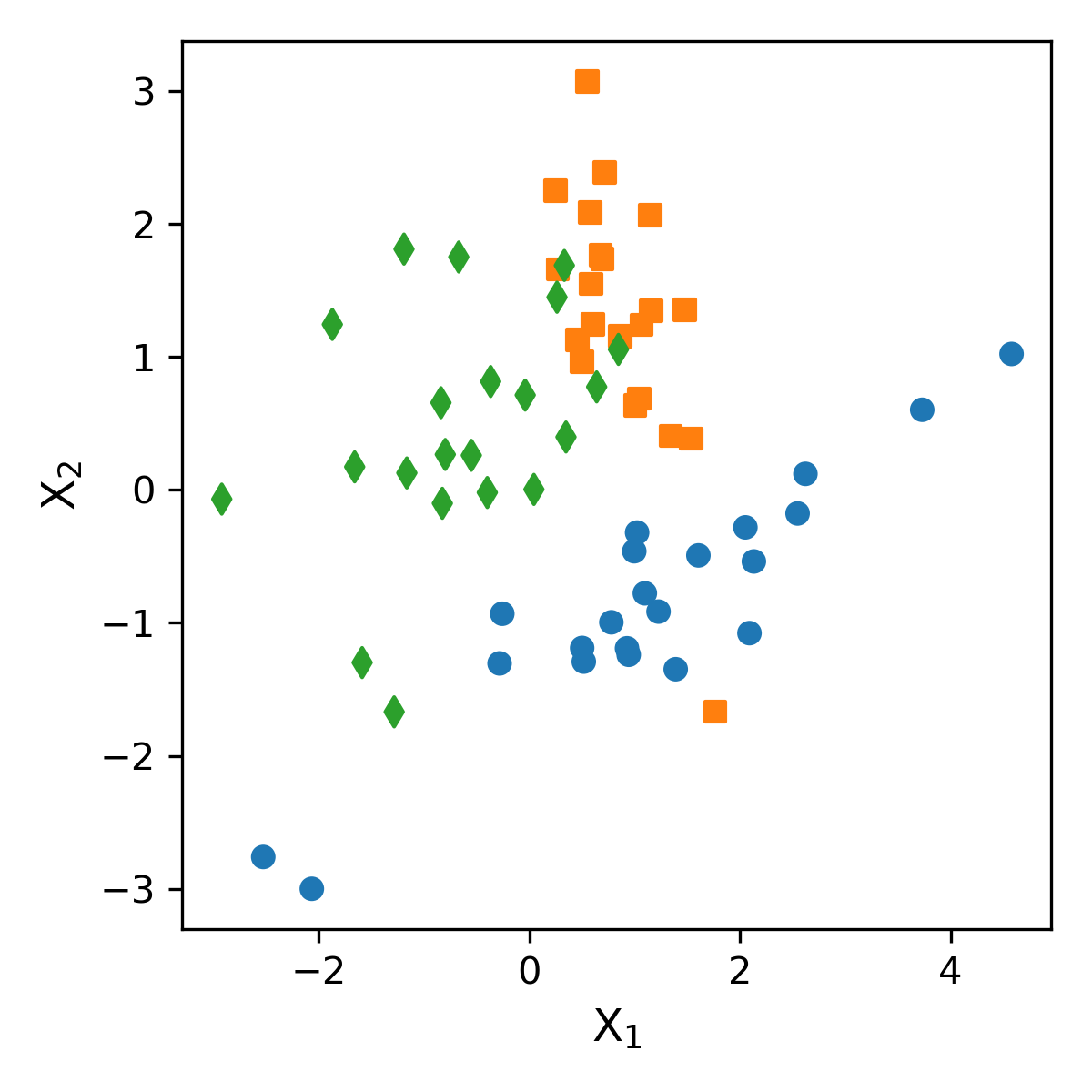}
 \caption{Data for clustering.  The different colors and shapes indicate class membership of the samples.}
\label{fig:scikit_inputs}
\end{figure}

For our example here, we use the \texttt{scikit-learn} function \texttt{make\_classification} to make a 2-dimensional, 3-class, 64-sample clustering problem.
The classes are approximately balanced with a single cluster per class.
The feature space is shown in Figure \ref{fig:scikit_inputs}.
We use 5 neural networks, each with a single hidden layer of 50 units with leaky ReLU activation \cite{xu_empirical_2015}.
All layers are fully-connected.
The output of the networks is a softmax function with three outputs.

Despite the low dimensionality of the example, we do not attempt to demonstrate superlative performance compared to, say, classical algorithms like DBSCAN \cite{schubert_dbscan_2017}, OPTICS \cite{ankerst_optics_1999}, or spectral clustering \cite{von_luxburg_spectral_2007}.
Rather, we aim to describe CoOL and how to use it, and in doing so we can introduce some definitions and concepts useful to understanding how it functions.

In this example we do not standardize the features of the input data.
This is because CoOL uses the structure of the data to find a partition of it and changing that structure by scaling the different dimensions is contra our goals.
See Figure 14.5 and the surrounding text of \cite{hastie_elements_2017} for a similar argument also in the context of clustering.

We note that this particular problem is actually quite difficult.
There are several green diamonds that occupy the space of the orange squares.
There is a single orange square separated from the others by the entire blue circle cluster.
There are also two blue squares in the lower left that are separated from the others by two green diamonds.
For these reasons, we do not expect any clustering algorithm to achieve 100\% accuracy.

The samples are fed into all 5 neural networks as a 64-sample batch.
A key requirement of CoOL is simultaneity of samples for all the networks, i.e. each network must be shown the same samples at the same time.
It is not possible to, for example, reorder the samples for one of the networks.
In this example, the samples shown to each network are identical and the networks are also identical (but not copies).
This is a convenience, the samples shown to each network need not be the same, they just need to be related by simultaneity.
We discuss this as further work in Section \ref{sec:conclusions}.

\begin{figure}
 \centering
        \includegraphics[width=0.5\textwidth]{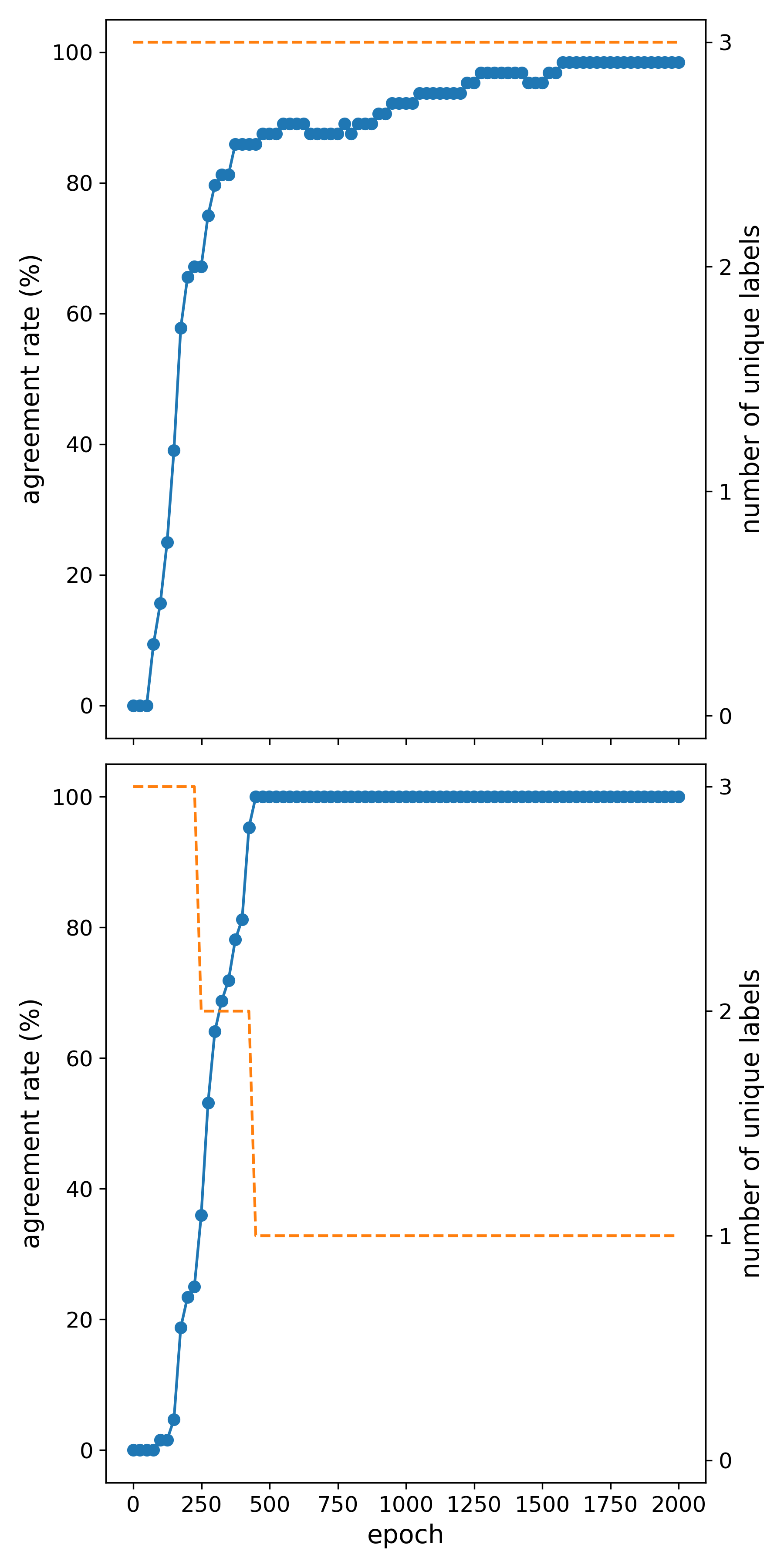}
 \caption{Agreement rate (blue, solid line and circles) and number of unique labels (orange, dashed line) during training on the clustering problem shown in Figure \ref{fig:scikit_inputs}. The top plot shows the case when learning is regularized by the determinants of the error rate matrices of the neural network outputs.  The bottom plot shows the case without this regularization.}
\label{fig:agreement_rate}
\end{figure}

The networks are trained for 2000 epochs with a learning rate of $10^{-4}$ and no weight decay using the Adam optimizer \cite{kingma_adam_2017}.
During training, we monitor two quantities: the agreement rate and the number of unique labels produced by CoOL.
Both of these are shown in Figure \ref{fig:agreement_rate}.
Agreement rate is the fraction of samples where all of the observers agree on which clusters the samples belong to.
The number of unique labels is the number of different labels that the networks produce for the samples that they all agree on.

\begin{figure}
 \centering
        \includegraphics[width=0.5\textwidth]{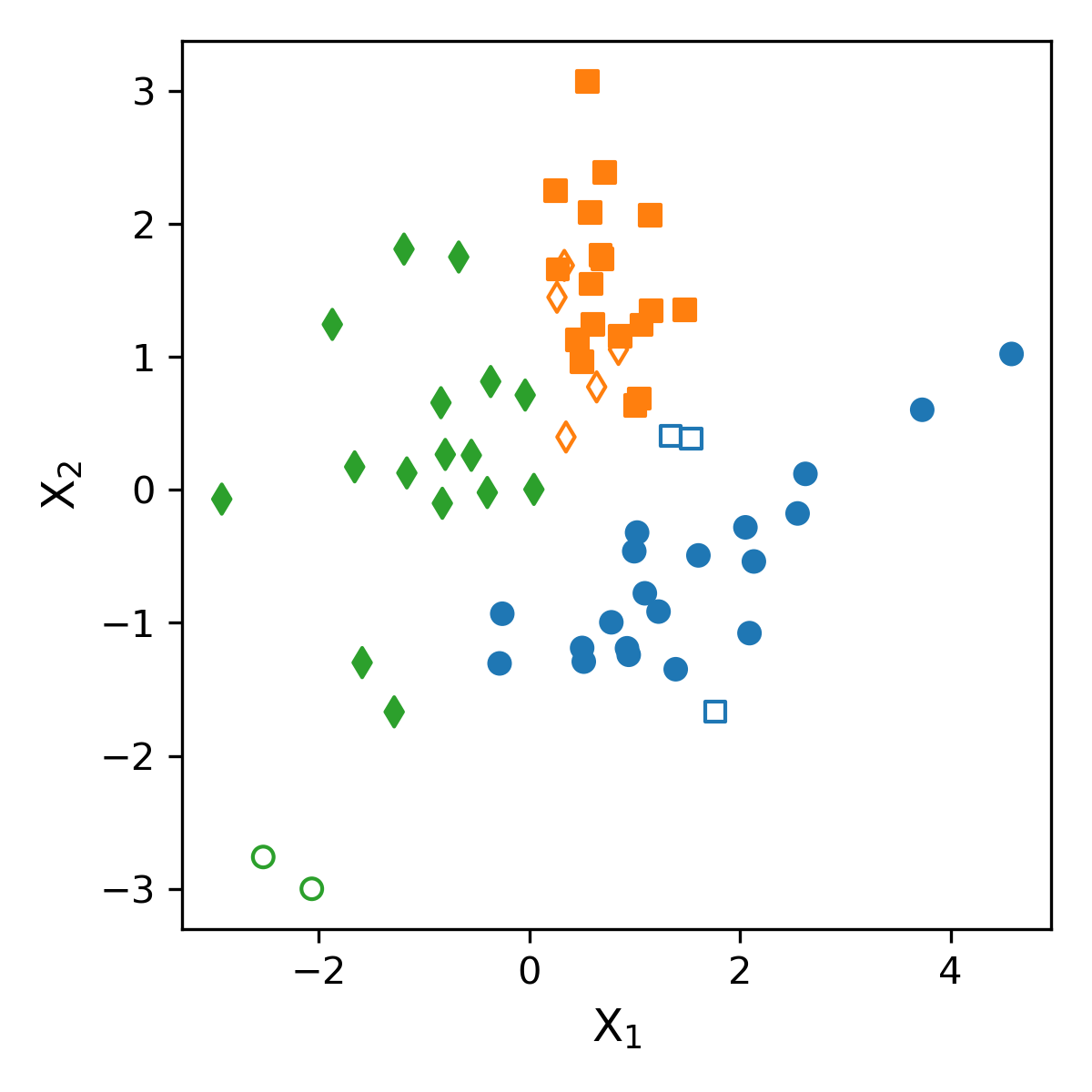}
 \caption{Results of clustering by CoOL.  The different colors and shapes indicate cluster membership of the samples.  The filled shapes show samples that were correctly clustered together while the unfilled shapes show samples that were incorrectly clustered together.  For the latter, the color represents the cluster CoOL assigned the sample to while the shape represents the true cluster identity.}
\label{fig:scikit_results}
\end{figure}

In this work we attempted to use the results of the EM-algorithm to estimate the ``true'' cluster for each sample and use cross-entropy to train the observers as if they were supervised.
We see in the bottom plot of Figure \ref{fig:agreement_rate}, that this was not successful.
The most expedient route for the observers to agree on the cluster labels for each sample is to agree that all of the samples belong to a single cluster.
This is both a strong attractor during optimization and not useful.
This degenerate case has an easy-to-identify signature in that the reliability matrix for each of the observers has only a single column with non-zero elements.
To regularize against this, we add a term to the loss proportional to the absolute value of the determinant of each of the observers.
See Sections \ref{sec:learning} and \ref{app:nearconvergence} for more details on this term.
As shown in the top plot of Figure \ref{fig:agreement_rate}, this regularization helps CoOL learn clusters for the samples.

The results of the clustering by CoOL are shown in Figure \ref{fig:scikit_results}.
As predicted due to the difficulty of the problem, some of the samples were incorrectly clustered.
However, we see that CoOL has done a reasonable job of clustering the samples.
The discrimination boundaries are clearly non-linear.
It also does not appear to have maximized the margins between the clusters \cite{cortes_support-vector_1995}; see the two blue empty squares to the lower right of the filled orange squares.
Nevertheless, CoOL has done a good job finding the clusters without any kind of distance- or similarity-based hyperparameter.

To generate labels using CoOL we show the samples to each of the networks to obtain a probability distribution over the possible output labels.
We then say that a network's assignment is the maximum probability label.
It is then possible to reconcile the labels among the networks for each sample using any of the strategies for solving the label correspondence problem given in Section {\ref{sec:relatedwork}}.
In practice and for the work here, we find that all the networks agree on the label, so such reconciliation is not necessary.
Because of this, we simply take the output of the first model as the predicted labels.
Note that these labels are not related to the ``true'' labels in the dataset, should any exist, except through the mapping made by CoOL.
It is possible for the number of output labels to be different from the number of true labels.
We cover this topic in Section {\ref{subsec:otherscenarios}}.

\section{Related work}
\label{sec:relatedwork}

One way to view CoOL is as a clustering algorithm.
While most algorithms, including deep-learning based algorithms, rely on some notion of distance (with DBSCAN as a classical example \cite{schubert_dbscan_2017}) or similarity/affinity between samples (with spectral clustering as a classical example \cite{von_luxburg_spectral_2007}), CoOL relies on the self-expressiveness of the data.
This idea is similar to the one described in \cite{elhamifar_sparse_2013}, but we do not assume anything about how the samples relate to each other or the subspaces (manifolds) they inhabit.
An expansive survey of deep clustering is available in \cite{zhou_comprehensive_2022}.

Distance-based methods have been extended to deep learning typically by using auto-encoders to transform the data into a lower dimensional latent space.  
This strategy is common in the deep clustering literature \cite{guo_improved_2017, gao_deep_2019, niu_spice_2022, li_neural_2022, peng_xai_2022, svirsky_interpretable_2024}.  
These methods require some definition of closeness, density, or similarity in the reduced dimensionality (latent) space. 

Deep learning similarity/affinity-based methods have also been implemented  \cite{lv_pseudo-supervised_2021, shaham_spectralnet_2018, yang_joint_2016}.  
These kinds of algorithms can scale poorly to large sample sizes in general because the affinity computations scale as the number of samples squared unless special steps are taken.
Recent strategies tend to focus on approximating the affinity computations to avoid the unappealing scaling.
Two examples are SpectralNet, which approximates the affinity matrix using mini-batch sampling \cite{shaham_spectralnet_2018}, and Deep Embedded Subspace Clustering, which treats the bases of the affinity function as parameters to be learned \cite{cai_efficient_2022}.
Others have used ensembles of networks, trained independently, to compute the similarity of samples through consensus of those networks \cite{gupta_unsupervised_2020}.

Another way to view CoOL is as a label reconciliation algorithm or ensemble algorithm for clustering.
A comprehensive survey of classical ensemble algorithms is given in \cite{vega-pons_survey_2011}, the most relevant are a series of works by Topchy \cite{topchy_combining_2003, topchy_mixture_2004, topchy_clustering_2005}.
Importantly, these works attempt to solve the label correspondence problem\footnote{The label correspondence problem is the task of associating labels from different labeling strategies with each other.} once a number of different models have produced cluster labels, whereas here we use learnable labels from a cohort of trainable observers to estimate a set of cluster labels.

The authors of \cite{greff_neural_2017} use expectation maximization to perform image segmentation.
Therein, they train multiple neural networks in parallel to simultaneously learn representations of image objects and the assignment of pixels to the different objects, which they call clusters.
EM has also been used to learn the parameters for stochastic neural networks and a mixture of experts \cite{ng_using_2004}, to train neural networks to learn a mixture model \cite{tissera_neural_2022, liu_neural_2020}, and to train a text encoder while simultaneously cleaning noisy labels \cite{chen_neural_2023}.


CoOL shares a passing similarity with methods that learn about data by comparing two unlabeled streams of data, such as deep canonical correlation analysis \cite{andrew_deep_2013}, maximum covariance analysis \cite{bretherton_intercomparison_1992} (and its deep version \cite{luo_vitac_2018}), and coincidence anomaly detection \cite{humble_coincident_2024}.

\section{Learning}
\label{sec:learning}

In this section we adapt the machinery of expectation maximization \cite{dempster_maximum_1977}, as described by Dawid and Skene \cite{dawid_maximum_1979} and shown diagrammatically in Figure \ref{fig:cool_diagram}, to the problem of unsupervised clustering.
We will discuss the process as batch-based, where each batch uses the whole dataset, and constitutes an epoch.
However, all of the formulae to come depend on a sum over the samples, so they can easily be used in a distributed data parallel fashion or in mini-batches.

We assume $K$ randomly initialized observers, which are any differentiable way to estimate the probabilities of $J$ possible classes.
We further assume that there are $I$ samples in our batch.
The output of observer $k$ for sample $i$ and class $j$ is given by $s_{ij}^{(k)}$, and we have $\sum_j s_{ij}^{(k)}=1$.
The first thing we need is an estimate for the probability of each sample being a member of each class, which we compute as the mean over all the observers,

\begin{equation}
    T_{ij}^{(0)} = \frac{1}{K}\sum_{k=1}^{K} s_{ij}^{(k)}.
    \label{eq:firstT}
\end{equation}

\noindent Where the superscript $(0)$ indicates that this is the first estimate.
Following Dawid and Skene, we can then compute the proportion of samples that belong to each class as 

\begin{equation}
    p_j = \frac{1}{I} \sum_{i=1}^{I} T_{ij}^{(t)}
    \label{eq:class_probs}
\end{equation}

\noindent for any iteration, $t$, of EM.  
We will suppress the superscript $(t)$ for all variables other than $T_{ij}^{(t)}$ because they are all estimated the same way given $T_{ij}^{(t)}$, whereas $T_{ij}^{(0)}$ is estimated differently from later values for $T_{ij}^{(t)}$.

To compute what we call the reliability matrices for each observer and remain consistent with the work of Dawid and Skene, we need a label for each sample made by each observer rather than a probability distribution over all possible labels.
To do that, we draw a single sample from the distribution represented by each observer's output for each sample.
In doing so, we create a binary matrix, $n_{ij}^{(k)}$, with a single 1 in each row, for each observer that is the same shape as Eq. \ref{eq:firstT}.
We can then compute the reliability matrix for each observer as

\begin{equation}
    \pi_{jl}^{(k)} = \frac{\tilde{\pi}_{jl}^{(k)}}{\sum_{l=1}^{J} \tilde{\pi}_{jl}^{(k)}}
    \label{eq:reliability}
\end{equation}

\noindent where we define 

\begin{equation}
    \tilde{\pi}_{jl}^{(k)} = \sum_{i=1}^{I} T_{ij}^{(t)} n_{il}^{(k)}
    \label{eq:unnorm_reliability}
\end{equation}

\noindent for any iteration, $t$, of EM.
We can then use Eq. 2.5 of \cite{dawid_maximum_1979} to compute a new estimate for $T_{ij}^{(1)}$ as

\begin{equation}
    T_{ij}^{(1)} = \frac{p_j \prod_{k=1}^{K} \prod_{l=1}^{J} \left( \pi_{jl}^{(k)} \right)^{n_{il}^{(k)}}}{\sum_{q=1}^{J} p_q \prod_{k=1}^{K} \prod_{l=1}^{J} \left( \pi_{ql}^{(k)} \right)^{n_{iq}^{(k)}}}
    \label{eq:t_ij_update}
\end{equation}

\noindent and repeat the process again.
The above formulae can be re-used for each iteration of EM.
The number of parameters to be estimated by EM is very small, $K J^2 + J$ per iteration.
This will require hundreds of iterations to reach the size of a small neural network, and here we focus on only a single iteration.
In other words, memory consumption is driven by gradient computation in the observers, not the EM parameters.

The goal for training is to minimize the negative log-likelihood function along with the regularization term

\begin{equation}
\label{eq:full_loss}
L = \sum_{k=1}^{K} L^{(k)} = \sum_{k=1}^{K} \left( -\alpha \left[ \sum\limits_{i=1}^{I} \sum\limits_{c=1}^{J} \mathbf{I}(y_{i} = c) \log(s_{ic}^{(k)})\right]  - \lambda |\det(\tilde{\pi}^{(k)})| \right).
\end{equation}

\noindent Where $\mathbf{I}(y_{i} = c)$ is the indicator variable and equals one when the label for sample $i$ is $c$ and zero otherwise.
The labels come from sampling $T_{ij}^{(t)}$ at the end of the EM process to produce a single label for each sample, similar to how $n_{ij}^{(k)}$ is generated for each observer.
We did this to achieve a model-driven balance between exploration of possible labels and exploitation of labels that networks agree on.
When all of the networks strongly agree on the sample label, that sample label is very likely to be chosen.
When the networks are indifferent to the label, any label can be chosen. 
This allows the algorithm to explore the label space.

The weight parameters, $\alpha$ and $\lambda$, allow us to change the importance of the two terms.
The first term in the parenthesis is the usual cross-entropy loss function for supervised learning and the second term is the regularization of the determinants of the observer's un-normalized reliability matrix.
We perform ablations studies over these parameters in Appendix {\ref{app:ablation}}.
We simply note here that $\lambda$ is a much more important parameter than $\alpha$ or the choice of the negative log-likelihood as the training function.

For an output $m$ of a particular observer $v$ on sample $h$ the derivative of the first term is given by 

\begin{equation}
    \left. \frac{\partial L}{\partial s_{hm}^{(v)}} \right|_{\lambda=0} = -\alpha \frac{\mathbf{I}(y_{h} = m)}{s_{hm}^{(v)}}.
    \label{eq:ce_grad}
\end{equation}

\noindent Thus, the gradient for most of the $s_{hm}^{(v)}$ are zero except for the outputs which have been selected through the sampling of $T_{ij}^{(t)}$.
In addition, the gradients are only large when a label is chosen for which $s_{hm}^{(v)}$ is small.
This matches our intuition about what should be happening during learning.
When all of the observers have long agreed on the label for a sample, their outputs are near 1 and the gradients are modest.
However, an observer which assigns a small probability to that label will see a correspondingly large gradient.
This disagreement dynamic also happens for the determinant term of Eq. \ref{eq:full_loss}, which we discuss in Appendix \ref{app:nearconvergence}.

The derivative of the second term is given by 

\begin{equation}
\begin{split}
    \left. \frac{\partial L}{\partial s_{hm}^{(v)}} \right|_{\alpha=0} & = -\lambda \sum_{k=1}^{K} \textrm{sign}(\det(\tilde{\pi}^{(k)})) \left[ \sum_{r,q=1}^{J} (-1)^{r+q} \det(\underline{\tilde{\pi}}_{rq}^{(k)}) \frac{\partial \tilde{\pi}_{rq}^{(k)}}{\partial s_{hm}^{(v)}} \right], \\
    & = -\lambda \sum_{k=1}^{K} \frac{\textrm{sign}(\det(\tilde{\pi}^{(k)}))}{K} \left[\sum_{q=1}^{J} (-1)^{m+q} \det(\underline{\tilde{\pi}}_{mq}^{(k)}) n_{hq}^{(k)} \right],
    \label{eq:pi_num_grad}
\end{split}
\end{equation}

\noindent where $\underline{\tilde{\pi}}_{mq}^{(k)}$ is the minor matrix of $\tilde{\pi}^{(k)}$ with row $m$ and column $q$ removed; the second line comes from combining Equations \ref{eq:firstT} and \ref{eq:unnorm_reliability} and substituting for $\tilde{\pi}_{rq}^{(k)}$ in the first line.
This gradient does not depend on $v$, the identity of the observer that produced the estimate $s_{hm}^{(v)}$, and sums over $k$, all of the observers.
Because of this, the gradient from this term is the same for all the observers.
In other words, the observers share the same gradient that is a kind of weighted average over the gradients of all the individual observers.
Unlike the ``selective'' gradient shown in Eq. {\ref{eq:ce_grad}}, and perhaps hard to tell, the gradient term in Eq. {\ref{eq:pi_num_grad}} is usually non-zero for all samples, $h$, and outputs, $m$.

This sharing of gradients may help get the observers out of local minima or saddle points.
As each observer is randomly initialized, it will start in a different part of the, potentially enormous, parameter space.
From there they share gradients and learn until they have agreed on the class label for most of the samples, i.e. the algorithm only converges when the parameters for all of the observers produce a nearly identical set of labels for all the samples.
Because the observers all start out in a different part of parameter space, we do not expect (and have not observed) them converging to the same location in parameter space.
Therefore, in order to agree on the label of the training samples, we expect that the observers will use the strongest features to partition the samples and that these partitions will generalize to samples not seen during training.
We provide evidence for this in Section \ref{subsec:clustermnist}.

We note that $\det(\tilde{\pi}^{(k)})$ is largest when the samples are equally partitioned among the classes.
This is not generally a problem, as shown in Section \ref{subsec:otherscenarios}, because we will use grouping to get more similarity information on the samples.

Why use $\tilde{\pi}_{rq}^{(k)}$ in Equation \ref{eq:pi_num_grad} instead of $\pi_{rq}^{(k)}$?
Taking the derivative of Equation \ref{eq:full_loss} with $\pi$ substituted for $\tilde{\pi}$, we find

\begin{equation}
    \left. \frac{\partial L}{\partial s_{hm}^{(v)}} \right|_{\alpha=0} = 
    -\lambda \sum_{k=1}^{K} \frac{\textrm{sign}(\det(\pi^{(k)}))}{K (\sum_{\ell=1}^{J} \tilde{\pi}_{m\ell}^{(k)})^2} \left[\sum_{q=1}^{J} (-1)^{m+q} \det(\underline{\pi}_{mq}^{(k)}) \sum_{i=1}^{I} T_{im}^{(0)} \left( n_{hq}^{(k)} - n_{iq}^{(k)}\right) \right],
    \label{eq:pi_grad}
\end{equation}

\noindent where we have made the same assumptions used in Equation \ref{eq:pi_num_grad}.
From the properties of determinants and using Equation \ref{eq:reliability} it is straightforward to see that $\textrm{sign}(\det(\pi^{(k)})) = \textrm{sign}(\det(\tilde{\pi}^{(k)}))$ and 

\begin{equation}
    \det{(\underline{\pi}_{mq}^{(k)})} = \frac{\det{(\underline{\tilde{\pi}}_{mq}^{(k)})}}{\prod_{w = 1,w \neq m}^{J} \sum_{u=1}^{J} \tilde{\pi}_{wu}^{(k)}}.
\end{equation}

\noindent  We note that the denominator in this equation is positive semi-definite and does not depend on $q$, so it can be moved outside the sum over $q$.
The terms $T_{im}^{(0)}$ are also positive semi-definite.
All of these facts combined mean that the term in in Equation \ref{eq:pi_grad} proportional to $n_{hq}^{(k)}$ always points in the same direction as Equation \ref{eq:pi_num_grad} and is always much smaller in magnitude.
However, the term proportional to $n_{iq}^{(k)}$ can be larger, and cause the gradients of Equations \ref{eq:pi_num_grad} and \ref{eq:pi_grad} to point in different directions.
We have found that Equation \ref{eq:pi_num_grad} leads to larger determinants of the $\pi^{(k)}$ matrices and much better partitioning of the samples.
A comparison of the evolution of the determinant of reliability for the two different regularization strategies is given in Figure \ref{fig:scikit_dets_and_seis}.

We leave further exploration of this facet of CoOL, including the effect of using more than one iteration of EM (implied by the use of Equation \ref{eq:firstT}), to future work.

\subsection{Understanding Convergence Without Labels}
\label{subsec:convergence}

In a typical clustering problem, we will not have access to labels or be able to see the clustering in low dimension as we did in Figure \ref{fig:scikit_results} to evaluate the progress of CoOL.
In order to determine the success of CoOL, defined as converging to an agreed-upon partition of most of the samples, we monitor five attributes: (1) loss, (2) agreement rate, (3) the number of unique classes, (4) the determinants of the $\pi^{(k)}$ matrices, and (5) the Shannon Equitability Index.
The loss tells us how well CoOL is learning to cluster the samples, lower is better and saturation indicates there is nothing left to learn.

We saw in Figure \ref{fig:agreement_rate} that agreement rate, the fraction of the samples on which all the observers agree on the label, grows while CoOL is learning, while the number of unique classes will drop if CoOL ends up in a degenerate solution.
Typically, on a successful run of CoOL, the number of unique classes never varies from the number defined by the user and the agreement rate gets well above 95\%.

\begin{figure}
 \centering
        \includegraphics[width=0.5\textwidth]{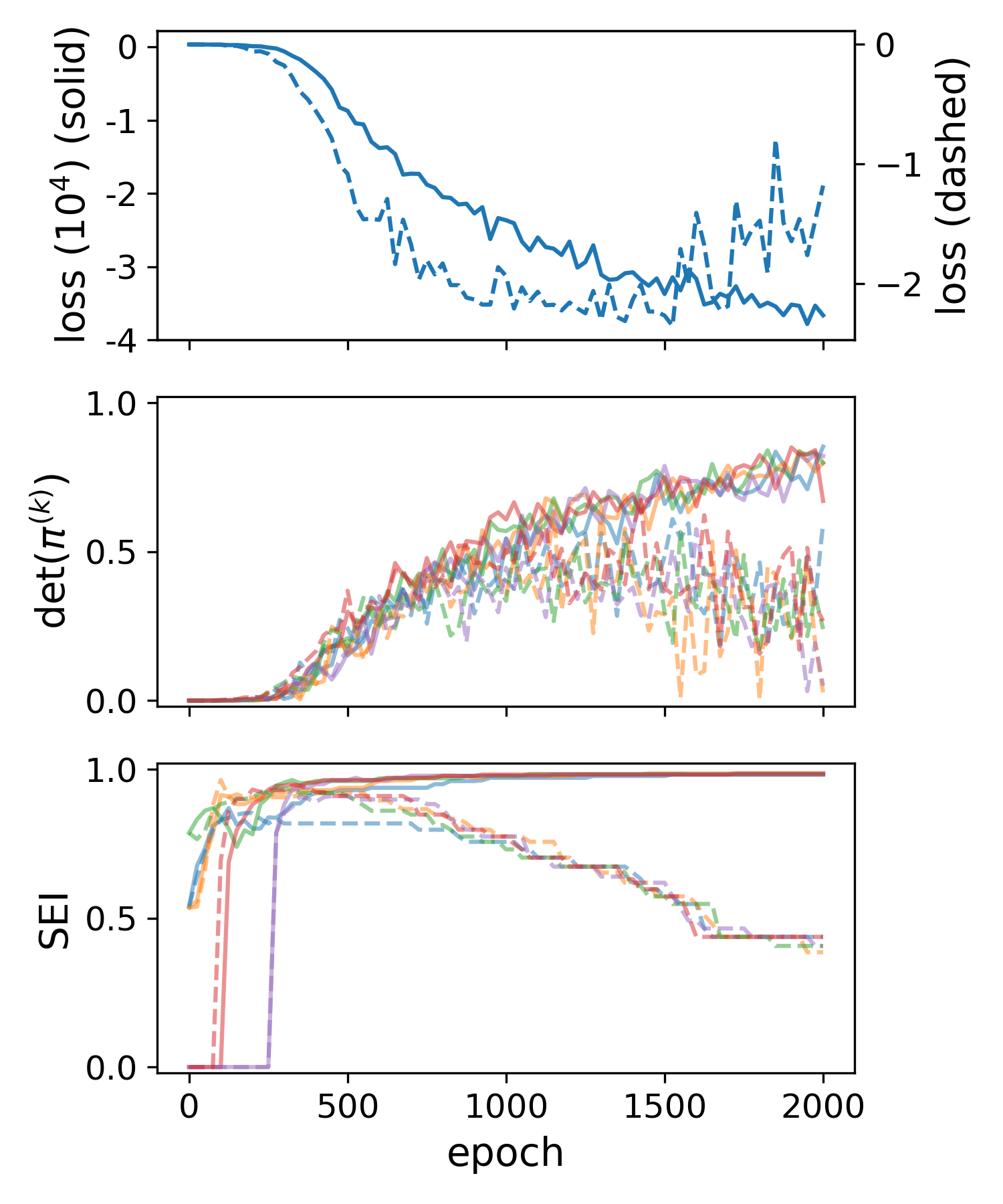}
 \caption{(top) Loss as function of epoch when learning the problem shown in Figure \ref{fig:scikit_inputs} using Equation \ref{eq:pi_num_grad} (solid lines) and Equation \ref{eq:pi_grad} (dashed lines).  (middle) $\det(\pi^{(k)})$ for the 5 observers trained. (bottom) Shannon Equitability Index (SEI) for the 5 observers trained.  Both runs resulted in agreement rate above 98\% and 3 classes.  However, the observers trained using Equation \ref{eq:pi_grad} put 48 of the 64 samples into a single cluster.}
\label{fig:scikit_dets_and_seis}
\end{figure}

In Figure \ref{fig:scikit_dets_and_seis} we see the evolution of the other attributes for a successful run (solid lines) and a run that was not successful (dashed lines).
It is clear from this figure that the determinant of the reliability matrix ($\pi^{(k)}$) grows for all the observers when the observers are learning to cluster the samples, but the same is not true for the less successful run.
The determinants might not get to one, even on a successful run, because a small number of samples in the ``wrong'' part of the reliability matrix can dramatically reduce the value of the determinant.
This problem is worse with a larger number of clusters.

The Shannon Equitability Index (SEI) is a measure of how equally each class is represented in a dataset, it is given by

\begin{equation}
    E_H = -\frac{\sum_{n=1}^{N} p_n \log(p_n)}{\log(N)},
\end{equation}

\noindent where $N$ is the number of classes and $p_n$ is the proportion of samples in class $n$.
$E_H$ takes on the value of 0 when there is only one class in the dataset and 1 when the dataset is equally partitioned between all the classes.
Equal partition of the samples is the most favored outcome of learning with Equation \ref{eq:full_loss}.
This is because such a partition maximizes the size of the $\tilde{\pi}^{(k)}$.

Between Figures \ref{fig:agreement_rate} and \ref{fig:scikit_dets_and_seis} the difference between a successful run of CoOL and several failure modes is clear.
In a successful run: the loss will continue to grow in magnitude; the agreement rate will approach unity; the number of unique classes will equal the number of allowed clusters; the determinant of the reliability matrices will grow and saturate; and the SEI will approach 1 for all observers.
The lack of any of these features is an indication that learning was not successful.
We reiterate that none of these metrics requires knowledge of the ``true'' cluster identity of the samples, if that information even exists.

\section{MNIST Example}
\label{sec:mnist}

Herein, we show CoOL clustering the MNIST digits.
We chose this example because the data is readily available from any number of sources and is typically used as a (nearly) first example for both deep learning packages and deep learning itself.
Throughout this section we will refer to problems by the number of classes that CoOL will have to cluster. 
The digits used during any experiment are counted up starting from zero to the number of classes less one.
As an example, when we say that an experiment uses 5 classes, we mean that it uses digits 0 through 4.
We will show both where CoOL succeeds and where it fails.
We emphasize here that we focus on image datasets because they are accessible, CoOL is not limited to working with images.

\subsection{Clustering MNIST digits}
\label{subsec:clustermnist}

\begin{figure}
 \centering
        \includegraphics[width=0.5\textwidth]{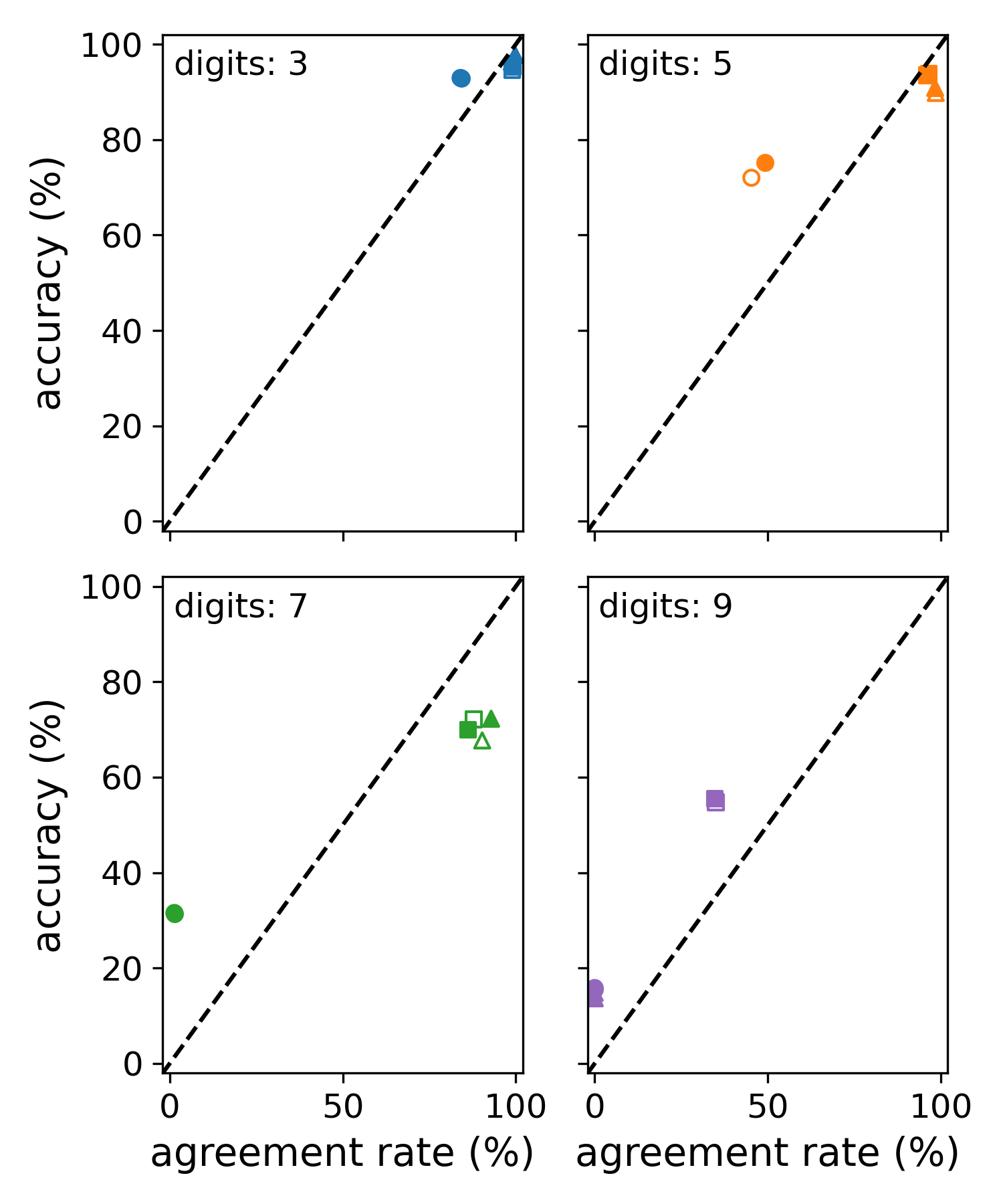}
 \caption{Plot of accuracy versus agreement rate for various hyperparameter settings on the MNIST digit problem.  The different markers indicate the learning rate used:  circles mean $10^{-5}$, squares mean $10^{-4}$, and triangles mean $10^{-3}$.  Filled markers indicate that a weight decay of $10^{-4}$ was used while empty markers mean no weight decay was used.  
 For this hyperparameter scan we used 288 samples per class.}
\label{fig:mnist_agr_rate}
\end{figure}

As is usually the case with any deep learning problem, the first thing we do is perform a hyperparameter scan.
In this scan, we vary the number of digits, learning rate, and weight decay factor, and keep $\alpha=\lambda=1$.
We train for 2000 epochs using the Adam optimizer and 288 samples per class.
Following Section \ref{subsec:convergence}, we judge the success of any given run by the agreement rate and the number of unique classes the observers use.
In Figure \ref{fig:mnist_agr_rate}, we compare the agreement rate with the accuracy of the clusters.
To compute accuracy, we assign to each cluster the true label that is most common in that cluster and then each sample in that cluster with that label is counted as a correct prediction while samples with a different label are counted as incorrect predictions.
We see that agreement rate (among the observers) is a fairly good predictor of accuracy, i.e. accuracy tends to be high when the agreement rate is high.
The exception is when the observers do not partition the samples into all of the desired classes, but this outcome is easily spotted, as shown in Section \ref{sec:description}.
It also appears that weight decay is not useful.
It may be the case that gradient sharing sufficiently guards against overfitting.

\begin{figure}
 \centering
        \includegraphics[width=0.5\textwidth]{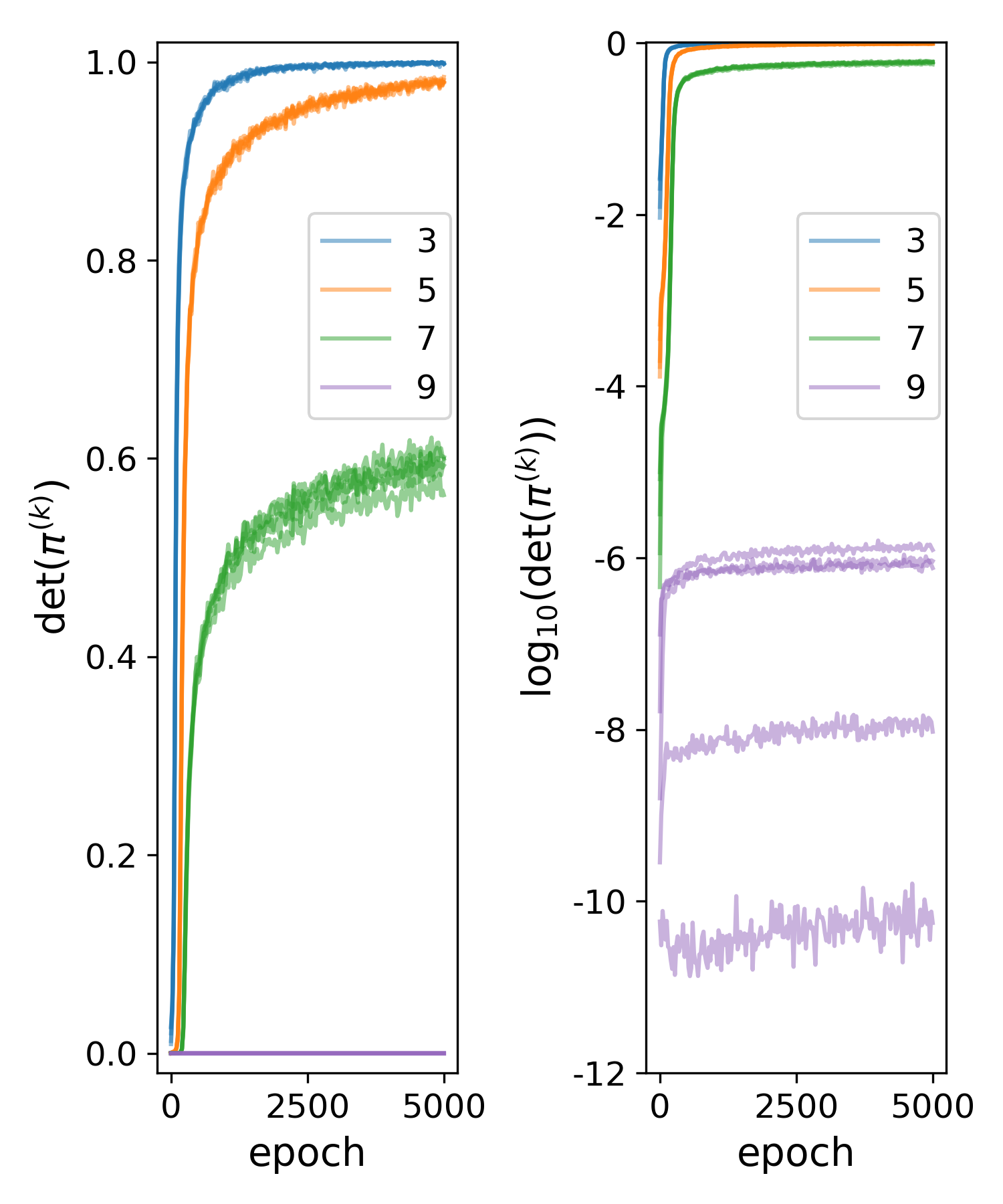}
 \caption{Determinant of the $\pi^{(k)}$ matrix for all 5 observers during select runs of CoOL on a varying number of digits with a varying number of classes (shown in the legends).  (left) shows a linear scale and (right) shows a log scale.  For completeness, these plots show training through 5000 epochs regardless of when early stopping occurred.}
\label{fig:mnist_det_pi}
\end{figure}

It is clear from Figure \ref{fig:mnist_agr_rate}, that the agreement rate (and the accuracy) both fall as the number of desired digits/classes increases.
In Figure \ref{fig:mnist_det_pi} we plot $|\det(\pi^{(k)})|$ (see Equation \ref{eq:reliability}) for each of the 5 observers during runs of CoOL with varying number of digits and classes.
From this point on, we train for 5000 epochs using 1000 samples per class.

The 9-class scenario is not converging because two of the observers are learning poorly (their determinants are stuck at or near zero), while the other three observers seem to learn for a little while.
The 7-class scenario shows determinants that are no longer growing at a value of about 0.6.
We will see in Appendix \ref{app:ablation} that CoOL performs better on the higher-class-count experiments with smaller values of $\lambda$.
It may be the case that a change in consensus label for a sample improves $|\pi^{(k)}|$ for some of the observers, but diminishes the same for other observers and this causes a contention.
The smaller shared gradients required for learning the higher-class-count clusters suggests that the region of convergence is smaller and learning weight decay strategies might be useful.
We leave investigation of this issue for future work and note that in Section \ref{subsub:wrongclusters} we discuss how we can use grouping to handle clustering that is not limited by the number of clusters assumed during an individual run.

In Table {\ref{tab:mnist_success}} we provide the agreement rate, accuracy, normalized mutual information (NMI) {\cite{Witten2005}}, and adjusted rand index (ARI) {\cite{hubert_comparing_1985}} at the end of training.
The latter three are computed on a hold-out set.
We stop training CoOL when the agreement rate reaches 99.5\% or epoch 5000 is reached, which ever is first.
All experiments are performed five times; we show the mean and standard deviation of all the quantities.
We train two different architectures using CoOL: one based on convolutional neural networks (CoOL-Conv) and another that uses a fully-connected network (CoOL-Enc).\footnote{CoOL-Enc uses the (not pre-trained) IDEC encoder architecture \cite{guo_improved_2017}.}
For comparison, we also show the performance of related algorithms: Improved Deep Embedded Clustering (IDEC) {\cite{guo_improved_2017}}, inTerpretable nEuraL cLustering (TELL) {\cite{peng_xai_2022}}, SpectralNet {\cite{shaham_spectralnet_2018}}, and k-Means {\cite{James_introduction_2013}}.
All of these algorithms are trained using the parameters and architectures that were used to train them on MNIST in their respective publications, except for k-Means, where the default parameters in \texttt{scikit-learn} are used {\cite{scikit-learn}}.
The algorithms are trained from scratch except for IDEC for which weights are available that were pre-trained on the entire MNIST dataset {\cite{xie_unsupervised_2016}}.

IDEC has the highest accuracy across all the datasets.
This is hardly surprising given the extensive pretraining of the autoencoder that was available.
Such pretraining might not be available in other contexts, as there might not be enough data, but IDEC serves as a point of comparison for the other methods.
The second highest accuracy for each of the cases belongs to CoOL-Conv (3-classes) and TELL (5-, 7-, and 9-classes).
CoOL-Conv is third in both the 5- and 7-class cases, performing nearly as well as TELL, while TELL is third in the 3-class case.
SpectralNet appears to be consistently miss-clustering the hold-out set because the NMI and ARI are fair to good, but the accuracy is not.
The performance of the fully-connected version of CoOL is modest.
It is interesting that the performance of K-Means is better as the number of classes is increased (in terms of all metrics) and the same is true for SpectralNet (in terms of NMI and ARI).
Because the datasets get larger with larger class counts, both because there are more classes and and there are more samples overall, this suggests that these algorithms benefit from more data and/or class diversity.
CoOL appears to do best when the class count is modest, which was foreshadowed in Figures {\ref{fig:mnist_agr_rate}} and {\ref{fig:mnist_det_pi}}.

For these runs, we set $\alpha=\lambda=1$ as suggested default parameters, in order to make a fair comparison with the other methods, for which we took the parameters used during learning on MNIST in their respective publications.
In Appendix {\ref{app:ablation}} we show that performance on the higher class-count problem can be dramatically improved by changing the $\lambda$ parameter while monitoring the agreement rate.
In addition, we discuss the computational performance of CoOL in Appendix {\ref{app:computation}}.

\begin{table}
\caption{
Comparison between different deep clustering models on the MNIST dataset.  The columns show the number of classes used from the MNIST dataset.
For each model, we show the accuracy on a hold-out set (acc.), normalized mutual information (NMI), and adjusted rand index (ARI).
In addition, the CoOL models show the agreement rate.
Each experiment was run 5 times to generate a mean and standard deviation as shown.
}
\centering
\begin{tabular}{| l l || c c c c |}
\hline
& & 3-MNIST & 5-MNIST & 7-MNIST & 9-MNIST \\
\hline
CoOL-Conv & agr. rate & 0.995$\pm$0.000 & 0.995$\pm$0.000 & 0.959$\pm$0.007 & 0.001$\pm$0.002 \\
          & acc. & 0.957$\pm$0.060 & 0.834$\pm$0.080 & 0.697$\pm$0.120 & 0.259$\pm$0.018 \\
          & NMI & 0.869$\pm$0.121 & 0.719$\pm$0.101 & 0.654$\pm$0.077 & 0.141$\pm$0.028 \\
          & ARI & 0.892$\pm$0.138 & 0.692$\pm$0.130 & 0.589$\pm$0.101 & 0.073$\pm$0.017 \\
\hline
IDEC & acc. & 0.989$\pm$0.001 & 0.965$\pm$0.001 & 0.885$\pm$0.001 & 0.827$\pm$0.001 \\
     & NMI & 0.941$\pm$0.004 & 0.901$\pm$0.002 & 0.842$\pm$0.001 & 0.764$\pm$0.001 \\
     & ARI & 0.967$\pm$0.003 & 0.916$\pm$0.002 & 0.800$\pm$0.001 & 0.699$\pm$0.001 \\
\hline
TELL & acc. & 0.874$\pm$0.013 & 0.864$\pm$0.018 & 0.752$\pm$0.020 & 0.617$\pm$0.014 \\
     & NMI & 0.711$\pm$0.017 & 0.695$\pm$0.028 & 0.576$\pm$0.023 & 0.509$\pm$0.019 \\
     & ARI & 0.684$\pm$0.024 & 0.692$\pm$0.032 & 0.541$\pm$0.032 & 0.420$\pm$0.021 \\
\hline
SpectralNet & acc. & 0.383$\pm$0.288 & 0.106$\pm$0.141 & 0.092$\pm$0.062 & 0.131$\pm$0.110 \\
            & NMI & 0.577$\pm$0.118 & 0.680$\pm$0.140 & 0.769$\pm$0.037 & 0.856$\pm$0.005 \\
            & ARI & 0.486$\pm$0.165 & 0.573$\pm$0.234 & 0.671$\pm$0.073 & 0.865$\pm$0.006 \\
\hline
CoOL-Enc & agr. rate & 0.997$\pm$0.001 & 0.996$\pm$0.001 & 0.645$\pm$0.440 & 0.046$\pm$0.048 \\
         & acc. & 0.817$\pm$0.116 & 0.658$\pm$0.125 & 0.436$\pm$0.165 & 0.238$\pm$0.112 \\
         & NMI & 0.604$\pm$0.174 & 0.507$\pm$0.111 & 0.336$\pm$0.189 & 0.124$\pm$0.114 \\
         & ARI & 0.610$\pm$0.203 & 0.452$\pm$0.150 & 0.253$\pm$0.142 & 0.077$\pm$0.074 \\
\hline
K-Means & acc. & 0.312$\pm$0.004 & 0.484$\pm$0.002 & 0.537$\pm$0.073 & 0.638$\pm$0.021 \\
        & NMI & 0.267$\pm$0.005 & 0.475$\pm$0.001 & 0.482$\pm$0.027 & 0.532$\pm$0.004 \\
        & ARI & 0.158$\pm$0.007 & 0.350$\pm$0.001 & 0.376$\pm$0.037 & 0.443$\pm$0.008 \\
\hline
\end{tabular}
\label{tab:mnist_success}
\end{table}

\subsection{Other Scenarios}
\label{subsec:otherscenarios}

In this section we demonstrate CoOL's performance beyond the somewhat idealized cases of the previous sections.

\subsubsection{The Wrong Number of Clusters}
\label{subsub:wrongclusters}

Up until this point and for convenience of the exposition, we have run CoOL on datasets where we knew the number of clusters ahead of time.
Typically, this information is not known ahead of time and we now turn to the matter of understanding the clustering without this information.
In Section \ref{subsec:convergence} we demonstrated a method for monitoring convergence without cluster labels, so the labels are not necessary to determine whether or not CoOL has converged to a partition of the samples.
There are still several outstanding questions: (1) how can we be assured that the clusters are not random and (2) how do we deal with the arbitrary labels given to the clusters by CoOL?
The crux of the problem of the second question is that on any two runs of CoOL, the label given to a sample need not be consistent.
For example, a sample might be put in cluster 0 for the first run, cluster 3 in the second run, and cluster 2 on the third run.
In fact, we need not even assume the same number of clusters for the different runs.

The solution begs the question about the nature of clustering.
Clustering only works, no matter the algorithm, if some set of the samples are more similar to each other than they are to some other set of samples.
Therefore, if CoOL is clustering in some sensible and consistent way, more-similar samples will be clustered together on each successive run of CoOL, even though the arbitrary label is changing for each run.
To understand the consistency of the clustering, we assign each sample to a ``group'' that is simply the concatenation of the cluster assignments given on the various runs, in order.
To continue the example sample described earlier, it would have group (0, 3, 2).

It is reasonable to ask, at this point, which cluster to use for a given run if all of the observers in that run do not agree on the cluster.
The easiest thing to do is to separate those samples from the others.
As we saw in Section \ref{sec:mnist}, we can train until the number of disagreements is small, so this is hardly a hindrance to the method.
We use this strategy here, but note that the fact that observers cannot agree on a sample's cluster is information that can be exploited by the user to investigate, for example, anomalies.

\begin{figure}
 \centering
        \includegraphics[width=0.5\textwidth]{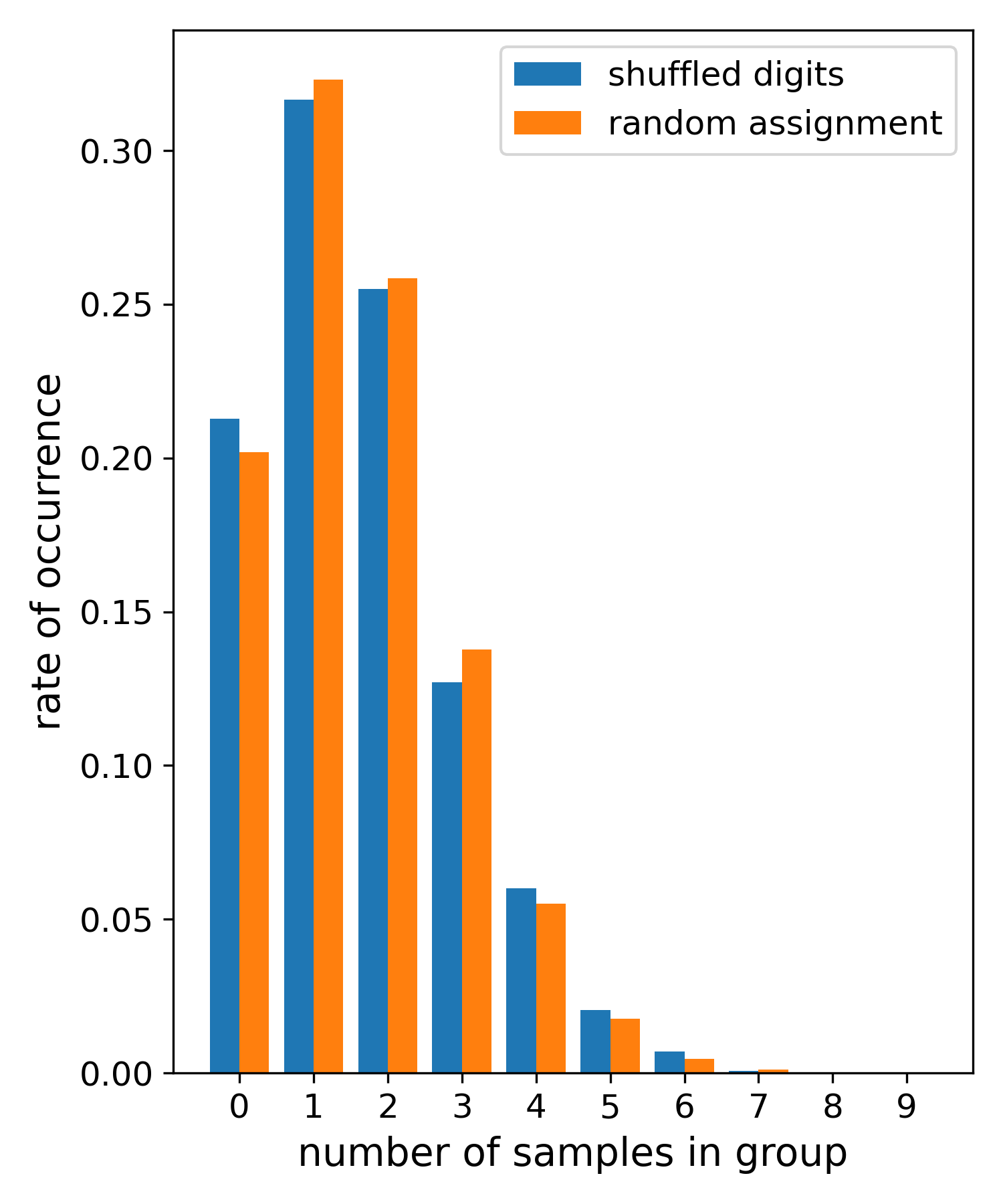}
 \caption{The fraction of groups containing the given number of samples when the features of the MNIST digits are shuffled on a per-sample basis.  The distribution is very close to random assignment of the samples to the groups.}
\label{fig:shuffled_features}
\end{figure}

With a group label for each sample, we can immediately answer the question as to random assignment.
If the assignments are random, then the group labels would also be random and the number of members in each group would follow a binomial distribution with mean given by $I / (\prod_{r=1}^{R}C_r)$, where $I$ is the number of samples, $C_r$ is the number of clusters used for run $r$, and $R$ is the number of runs.
In Figure \ref{fig:shuffled_features} we show the distribution of the number of samples per group when we shuffle the features of each sample.
The distribution is quite similar to random assignment.

\begin{table}
\caption{Table of all the groups with at least 10 samples found when clustering 5 digits into 4 classes using CoOL.  The grouping produced 64 groups total.  Group is the group name for a set of samples; count is the number of samples in that group; label is the most common label in the group; labels is a list of all the labels found in the group; and consistency is the fraction of the group samples that have the label given in the label column.  A consistency of 1 means that all of the group samples have the same label.}
\centering
\begin{tabular}{l r c c c}
\hline
Group & count & label & labels & consistency \\
\hline
(1, 3, 3, 3, 3) & 877 & 0 & (0)       	& 1.000 \\
(3, 2, 0, 2, 0) & 774 & 4 & (0 2 3 4) 	& 0.995 \\
(0, 0, 2, 1, 2) & 764 & 1 & (1 2)     	& 0.995 \\
(2, 1, 1, 0, 1) & 411 & 3 & (2 3)     	& 0.968 \\
(2, 1, 0, 0, 0) & 234 & 2 & (2)       	& 1.000 \\
(2, 1, 1, 1, 2) & 135 & 2 & (1 2 3)   	& 0.948 \\
(1, 1, 1, 0, 1) & 103 & 3 & (2 3)     	& 0.990 \\
(0, 0, 2, 2, 2) &  79 & 1 & (1)       	& 1.000 \\
(1, 3, 1, 0, 1) &  50 & 3 & (3)       	& 1.000 \\
(2, 2, 0, 0, 0) &  44 & 2 & (2)       	& 1.000 \\
(3, 2, 0, 0, 0) &  15 & 2 & (2)       	& 1.000 \\
(2, 1, 1, 0, 2) &  14 & 2 & (2 3)     	& 0.857 \\
(3, 1, 1, 0, 1) &  14 & 3 & (3)       	& 1.000 \\
(2, 1, 0, 0, 2) &  11 & 2 & (2)       	& 1.000 \\
(2, 3, 0, 0, 0) &  10 & 2 & (2)       	& 1.000 \\
(3, 1, 1, 2, 1) &  10 & 3 & (3)       	& 1.000 \\
\hline
\end{tabular}
\label{tab:groups_1}
\end{table}

To demonstrate the grouping process we ran CoOL 5 times.  The hyperparameters were the same for each run: 5 observers, 4 clusters, 5 digits, 1000 samples per digit, 2000 epochs, learning rate of $10^{-4}$, no weight decay, and the Adam optimizer.
We then group the samples as just described and show the ten largest groups in Table \ref{tab:groups_1}.
We see immediately that the groups are not random as we would expect, in that case, nearly 10 samples per group distributed over 1024 groups.
Instead, there are only 64 groups that contain any samples at all.

We see that CoOL has easily separated nearly 90\% of the zeros from the rest of the samples.
Performance on digits 1 and 4 is modestly inferior.
In contrast, CoOL is regularly clustering digits 2 and 3 together into smaller groups instead of two separate larger groups.
These group names are themselves meaningful as we would expect that samples from two group names with many matching digits will be more similar to each other than with a sample from a group with a totally different group name.
For example, the groups (2, 2, 0, 0, 0) and (3, 2, 0, 0, 0) are both small groups of digit 2.
The fact that several groups of digits 2 and 3 begin with the sequence (2, 1) or, even, (2, 1, 1) suggests that those digits look similar to CoOL.

The previous examples assume the number of clusters is very close to the true number of clusters and, under this constraint, we found that the clusters or groups typically represent digits.
In Appendix \ref{app:mnist_groups} we investigate what happens when the true number of clusters is much larger than the number of clusters assumed by CoOL.
We find that CoOL does not focus on digits, but rather large scale features of digits.
The largest group appears to be narrow digits that slope from the lower left to the upper right.
The second largest group appears to be digits with large loops.
The third and fourth groups, which are very close together in group-space being labeled (1, 0, 2, 3, 1) and (0, 0, 2, 3, 1), look both quite similar as boxes with tails down to the lower right and distinct in the patterns which fill in the legs of the box.
All of this is to say that CoOL has found a way to cluster and group 10 digits using only 4 clusters and those groups look reasonably consistent both within those groups and with nearby groups (and contrastive to more distant groups).

We now return to the two questions we asked at the beginning of this section.
The cluster assignments from CoOL do not appear to be random, which answers our first question.
We have also provided a method, i.e. grouping, for understanding how consistently CoOL clusters similar samples together.

Grouping provides a framework for a user to understand the cluster assignments in the data CoOL is trained on, but CoOL can give a cluster assignment for samples it was not trained on \cite{hamilton_inductive_2018}, we turn to what happens in that case next.

\subsubsection{Unseen Digits}
\label{subsub:unseendigits}

\begin{figure}
 \centering
        \includegraphics[width=0.5\textwidth]{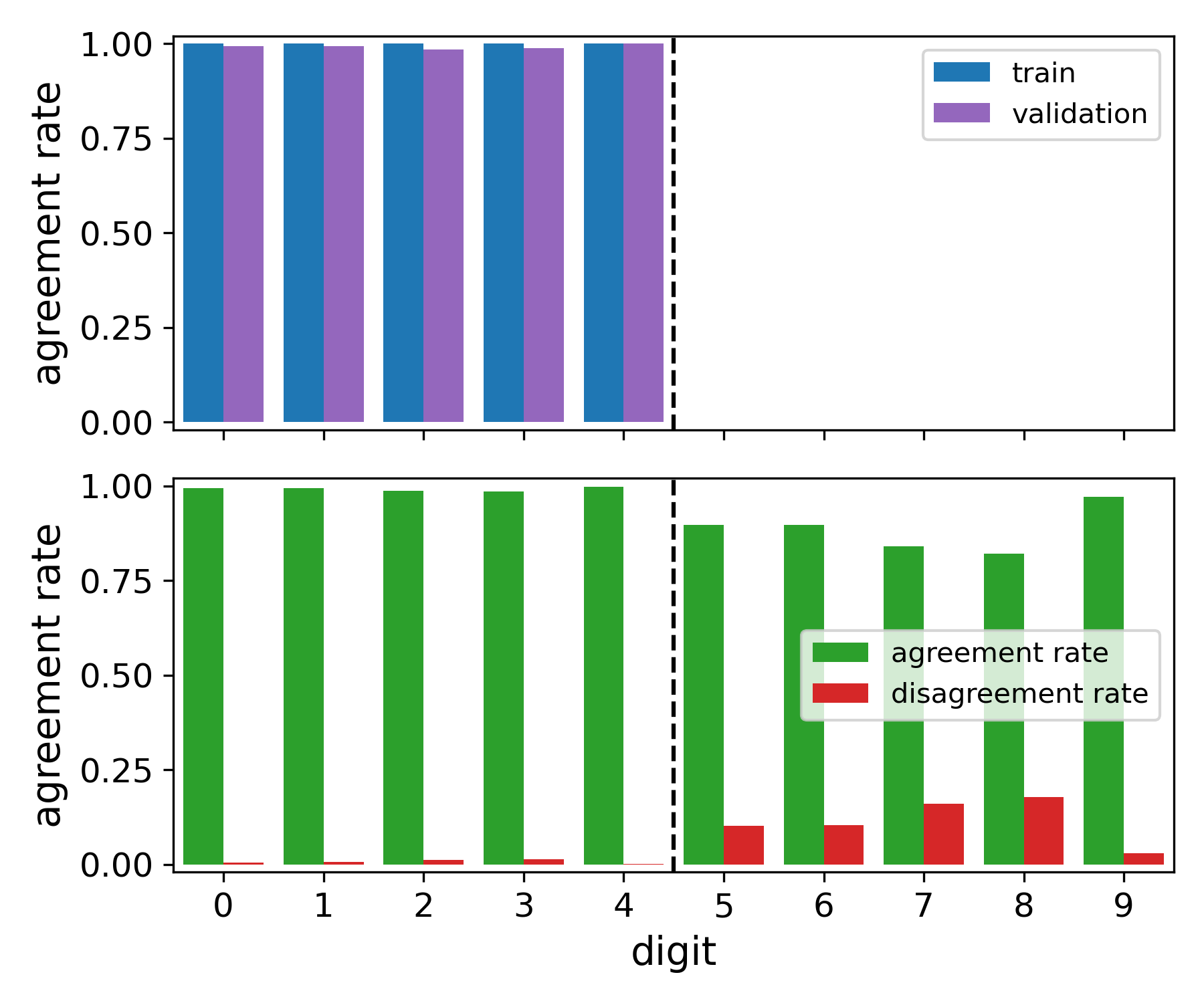}
 \caption{(top) The agreement rate for all digits (and samples) CoOL was trained or validated on.  (bottom) The agreement and disagreement rate for samples and digits not used during training.  The digits to the left of the dashed vertical line were used for training.}
\label{fig:unseen_digits}
\end{figure}

In Section \ref{subsec:clustermnist} we saw that CoOL performs similarly as well on a hold-out set of samples of the digits it was trained on.
This can be seen in a different way in Figure \ref{fig:unseen_digits}, where we see that the clustering of samples not used for training or validation is quite good (these samples are from digits that were used during training).
The bottom plot of Figure \ref{fig:unseen_digits} shows what happens when CoOL is shown samples from digits it was not trained on.

It is clear that digits that were not learned have much higher disagreement rate than those that were in the training set.
This demonstrates another useful feature of CoOL.
Because the clustering generalizes well, CoOL can be used to detect distributional drift of the input data.
In other words, CoOL can be used to detect changes in the data over time without understanding anything about the clusters.
This might be used to detect when new data is out of distribution (OOD) from a training set \cite{tagasovska_single-model_2019}. 
A recent example of this problem in the context of radiation management in a super-conducting accelerator can be found in \cite{goldenberg_data-driven_2025}.

As a hypothetical distributional shift, suppose we monitor the disagreement rate as a function of time during a process that we have recently trained CoOL to cluster.
The disagreement rate is given by

\begin{equation}
    \Delta \doteq \frac{t_d + u_d}{t + u} = \delta_t \frac{1 + \frac{\delta_u}{\delta_t} \frac{u}{t}}{1 + \frac{u}{t}}.
    \label{eq:disrate_1}
\end{equation}

\noindent Where $t$ is the number of train-like samples and $t_d$ is the number of train-like samples that CoOL disagrees on; $u$ and $u_d$ are defined similarly for the ``unlike'' samples, i.e. those samples unlike the ones seen during training, and $\delta_t = t_d / t$ ($\delta_u = u_d / u$) is the fraction of train-like (unlike) samples that are disagreed upon.
The ratio of unlike samples to train-like samples is

\begin{equation}
    \frac{u}{t} = \frac{\Delta - \delta_t}{\delta_u - \Delta}.
    \label{eq:disrate_2}
\end{equation}

If we assume, as seems reasonable from Figure \ref{fig:unseen_digits}, that $\delta_t=0.01$ and $\delta_u=0.1$, and take $\Delta = 2 \delta_t$, we find $u/t=$~1/8.
Thus, a modest number of OOD samples will cause a doubling in the overall disagreement rate if the new samples are well out of distribution from the training samples.
Further exploration of this topic is context dependent and is the subject of future work.


\section{Conclusions and Future Work}
\label{sec:conclusions}

We have described CoOL, a method for clustering data while simultaneously training differentiable models to do the same with end-to-end learning.
CoOL can cluster data with any dimensionality compatible with the models.
On the experiments we ran, CoOL returned reasonable clusters.
Along the way, we gave formulae for the gradients passed back to the models and a method for monitoring convergence without ground truth labels.
We have also shown that, even in the absence of interpretable clusters, a trained set of models can be used to detect distributional drift by monitoring the disagreement rate between the models.

Throughout this work, we have noted several areas for future work.
The question on what is the largest number of clusters that can be used with CoOL remains open, as does how many iterations of EM can be used productively.
In addition, different methods for selecting $n_{il}^{(k)}$ are an open topic of research, especially less stochastic ones.
We anticipate that there are other, potentially superior, ways to regularize learning.
For example, the entropy of the partition of the samples should have similar properties as the determinant of the reliability matrix \cite{shannon_communication_1948}\hl{, maximizing the normalized mutual information might also offer some benefits {\cite{Witten2005}}}.
We further expect that having a relatively small number of labeled samples will allow semi-supervised training of the networks.

Another task is to use simultaneity to cluster different representations of the same set of concepts.
For example, three of the observers could be shown images while 2 others are shown sounds related to those pictures.
In this example, clustering in one input type might be easier than the other, and so simultaneity might be used to improve clustering in a difficult space, or simply to associate related sounds and images with the same cluster labels without supervision.
This is possible because CoOL can be used on any data type with an associated differentiable model.
Of course, we need not limit the number of input types to two or to images and sound.

\section{Appendices}


\subsection{Properties of the Determinant Near Convergence}
\label{app:nearconvergence}

Assume that we have $K$ observers that agree on $I-1$ of the samples, but there is one observer, $K$, that estimates a small probability that sample $I$ is a different class from all the other observers' estimates.  Further assume that these observers classify a fraction, $\beta$, of the $I-1$ samples as class 1 and $1-\beta$ as class 0.  All the observers classify sample $I$ as certainly class 0 except observer $K$, which assigns probably $\alpha$ that this sample is class 1. 

Under these assumptions, we find that $\hat{T}^{(0)}_{ij} = \delta_{j,c_i}$ for the first $I-1$ samples and $\hat{T}^{(0)}_{I0}=1 - \alpha/K$ ($\hat{T}^{(0)}_{I1}= \alpha/K$) for the remaining sample.  At this point, it is straightforward to compute the EM parameters.  First, we find $I p_0 = (1-\beta) (I-1) + 1 - \alpha/K$ and $I p_1 = \beta (I-1) + \alpha/K$.  For the $K - 1$ certain observers ($k \in \{1,  \ldots, K-1\}$), the un-normalized reliability matrix is 

\begin{equation}
\label{eq: sure_pi}
\tilde{\pi}^{(k)} =
\begin{pmatrix}
(1 - \beta) (I - 1) + 1 - \alpha/K & 0\\
\alpha/K & \beta (I - 1)
\end{pmatrix}
.
\end{equation}

For the uncertain observer, there are two possibilities.  If the sampling of the observer's outputs returns class zero ($n_{I0}^{(K)}=1$ and $n_{I1}^{(K)}=0$), then the reliability matrix is identical to that shown in Eq. \ref{eq: sure_pi}.  On the other hand, if that sampling returns class 1 ($n_{I0}^{(K)}=0$ and $n_{I1}^{(K)}=1$), the reliability matrix is

\begin{equation}
\label{eq: unsure_pi}
\tilde{\pi}^{(K)} =
\begin{pmatrix}
(1 - \beta) (I - 1) & 1 - \alpha/K\\
0 & \beta (I - 1) + \alpha/K
\end{pmatrix}
.
\end{equation}

In either of these matrices, and their normalized forms, there are only two free parameters, $\beta$ and $\alpha$, the latter of which represents observer $K$'s uncertainty about which class sample $I$ belongs to.

We take the derivative of the determinants of Eqs. \ref{eq: sure_pi} and \ref{eq: unsure_pi} with respect to an arbitrary parameter of one of the observers which we will call $x_r$ where the subscript $r$ references the observer number.  We note that the parameter $\alpha$ depends only on observer $K$, while $\beta$ depends on all of the observers.  For Eq. \ref{eq: sure_pi} the derivative is

\begin{equation}
\label{eq: sure_under}
\frac{\partial}{\partial x_r} (\det \tilde{\pi}^{(k)}) = (I - 1) \left[ (1 - 2 \beta) (I - 1) + 1 - \frac{\alpha}{K} \right] \frac{\partial \beta}{\partial x_r} - \frac{\beta (I - 1)}{K} \frac{\partial \alpha}{\partial x_r} \delta_{rK}
\end{equation}

\noindent and the derivative of Eq. \ref{eq: unsure_pi} is

\begin{equation}
\label{eq: unsure_under}
\frac{\partial}{\partial x_r} (\det \tilde{\pi}^{(K)}) = (I - 1) \left[ (1 - 2 \beta) (I - 1) - \frac{\alpha}{K} \right] \frac{\partial \beta}{\partial x_r} + \frac{(1 - \beta) (I - 1)}{K} \frac{\partial \alpha}{\partial x_r} \delta_{rK}.
\end{equation}

\noindent We can see that the determinant acts as expected on the $\alpha$ term.  When the sampling of observer $K$'s outputs chooses class 0, maximizing the determinant decreases $\alpha$.  However, when the sampling of observer $K$'s outputs chooses class 1, maximizing the determinant increases $\alpha$.  The expected coefficient for the $\partial \alpha / \partial x_r$ term is

\begin{equation}
\label{eq: ev_unnorm}
\alpha \frac{(1 - \beta) (I - 1)}{K} - (1 - \alpha) \frac{\beta (I - 1)}{K} = \frac{I - 1}{K} (\alpha - \beta).
\end{equation}

\noindent This expectation tells us that when $\alpha$ is greater than $\beta$ the former tends to increase.
In other words, when observer $K$ assigns a probability ($\alpha$) that sample $I$ is class 1 that is larger than the fraction of samples that all observers agree are class 1 ($\beta$), that probability tends to increase.

We now normalize the rows of Eqs. \ref{eq: sure_pi} and \ref{eq: unsure_pi} such that they sum to 1 before computing the derivative.  For Eq. \ref{eq: sure_pi} this yields

\begin{equation}
\label{eq: sure_der}
\frac{\partial}{\partial x_r} (\det \pi^{(k)}) = \frac{I - 1}{K I^2 p_1^2} \left[ \alpha \frac{\partial \beta}{\partial x_r} - \beta \frac{\partial \alpha}{\partial x_r} \delta_{rK} \right]
\end{equation}

\noindent while for Eq. \ref{eq: unsure_pi} this yields

\begin{equation}
\label{eq: unsure_der}
\frac{\partial}{\partial x_r} (\det \pi^{(K)}) = \frac{I - 1}{I^2 p_0^2} \left[ - (1 - \frac{\alpha}{K}) \frac{\partial \beta}{\partial x_r} + \frac{(1 - \beta)}{K} \frac{\partial \alpha}{\partial x_r} \delta_{rK} \right].
\end{equation}

\noindent These equations have similar properties, with respect to the $\partial \alpha / \partial x_r$ term, as the un-normalized case.  However, the expected coefficient of that term is not as easy to parse.  We find this expectation value is

\begin{equation}
\label{eq: ev_norm}
\frac{I - 1}{K I^2 p_0^2 p_1^2} \left[ \alpha (1 - \beta) p_1^2 - (1 - \alpha) \beta p_0^2 \right].
\end{equation}

To simplify this equation, we assume $\beta = 1/2$, in which case $p_0 \approx p_1 \approx 1/2$ as well.  
This yields an expectation that is $4 / I^2$ times smaller than Eq. \ref{eq: ev_unnorm} under the same assumptions.
The further motivates our use of the un-normalized reliability matrix in Section \ref{sec:learning}.


\subsection{Computational Performance of CoOL}
\label{app:computation}

Here we investigate the computational performance of CoOL.
We train the convolutional version of CoOL on an NVidia A100 GPU and record the time it takes to reach 99.5\% agreement rate on the validation set or 5000 epochs, which ever comes first.
We measure the memory reserved using the pytorch function \texttt{max\_memory\_reserved} that returns the maximum GPU memory managed by the caching allocator during training \cite{pytorch2017}.
In addition to the batch-learning version shown in Section \ref{sec:mnist}, we also train a mini-batch version where each epoch is split into 1000 sample batches.
Every 25 epochs we compute the accuracy, NMI, and ARI on both the training and validation set.
We fix the number of classes, $J$, at 5.
We vary the number of networks, $K$, between the values 3, 5, 7, and 9.
We also vary the number of samples per class, $I/J$, between the values 100, 300, 1000, 3000.
Each experiment is repeated 3 times.

\begin{figure}
 \centering
        \includegraphics[width=0.5\textwidth]{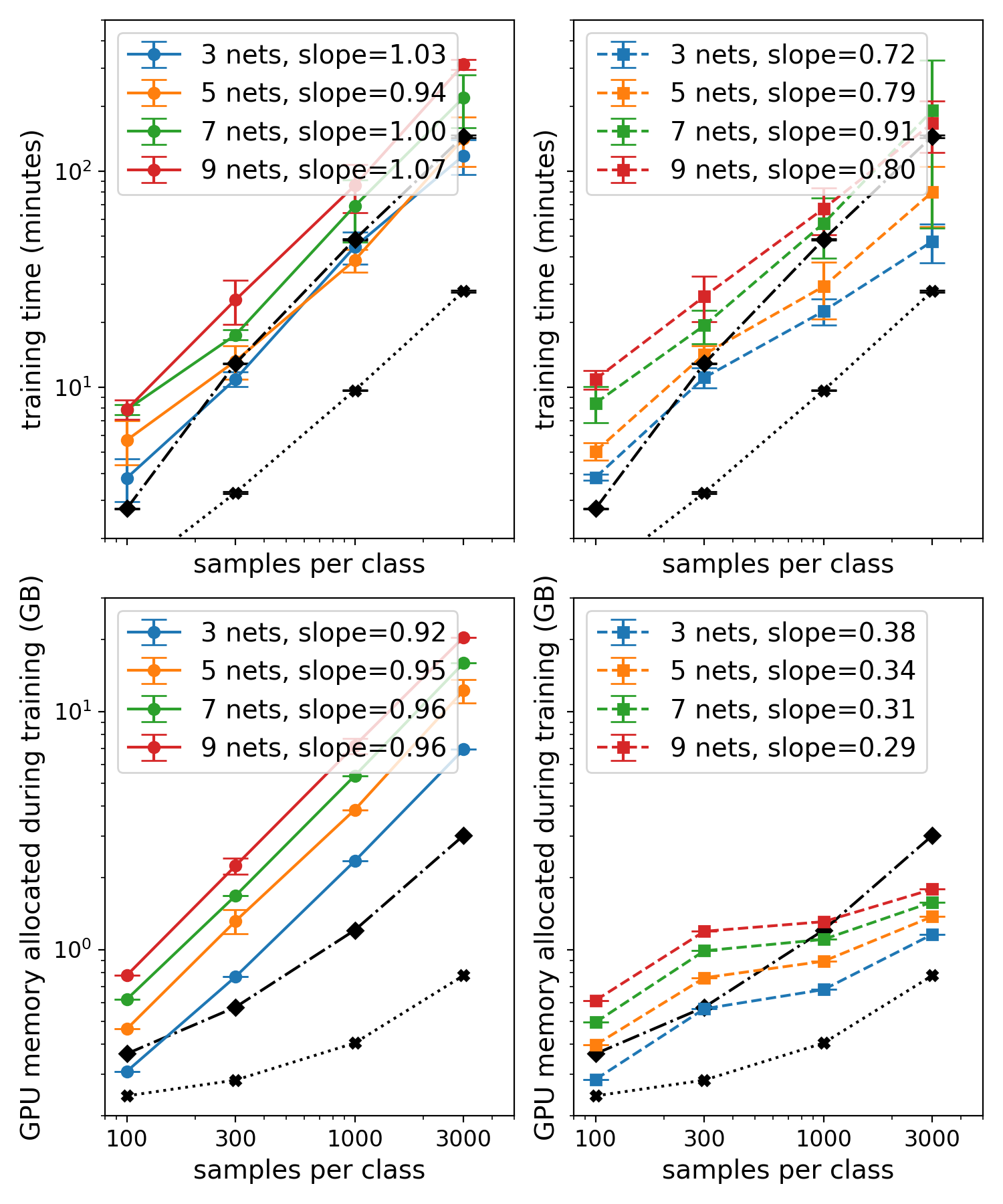}
 \caption{Statistics for training time and memory consumption when training CoOL with convolutional layers.  In the top row we show time to complete training and in the bottom row we show memory allocated by pytorch during training \cite{pytorch2017}.  The left column shows batch training and the right column shows mini-batch training.  The colored curves indicate the number of networks, $K$, used to learn 5 classes of the MNIST dataset as indicated in the legends, where the slope of the curves is also given.  The black dash-dot lines with diamond markers shows the performance of TELL and the black dotted lines with x markers shows the performance of SpectralNet.}
\label{fig:computation}
\end{figure}

\begin{figure}
 \centering
        \includegraphics[width=0.5\textwidth]{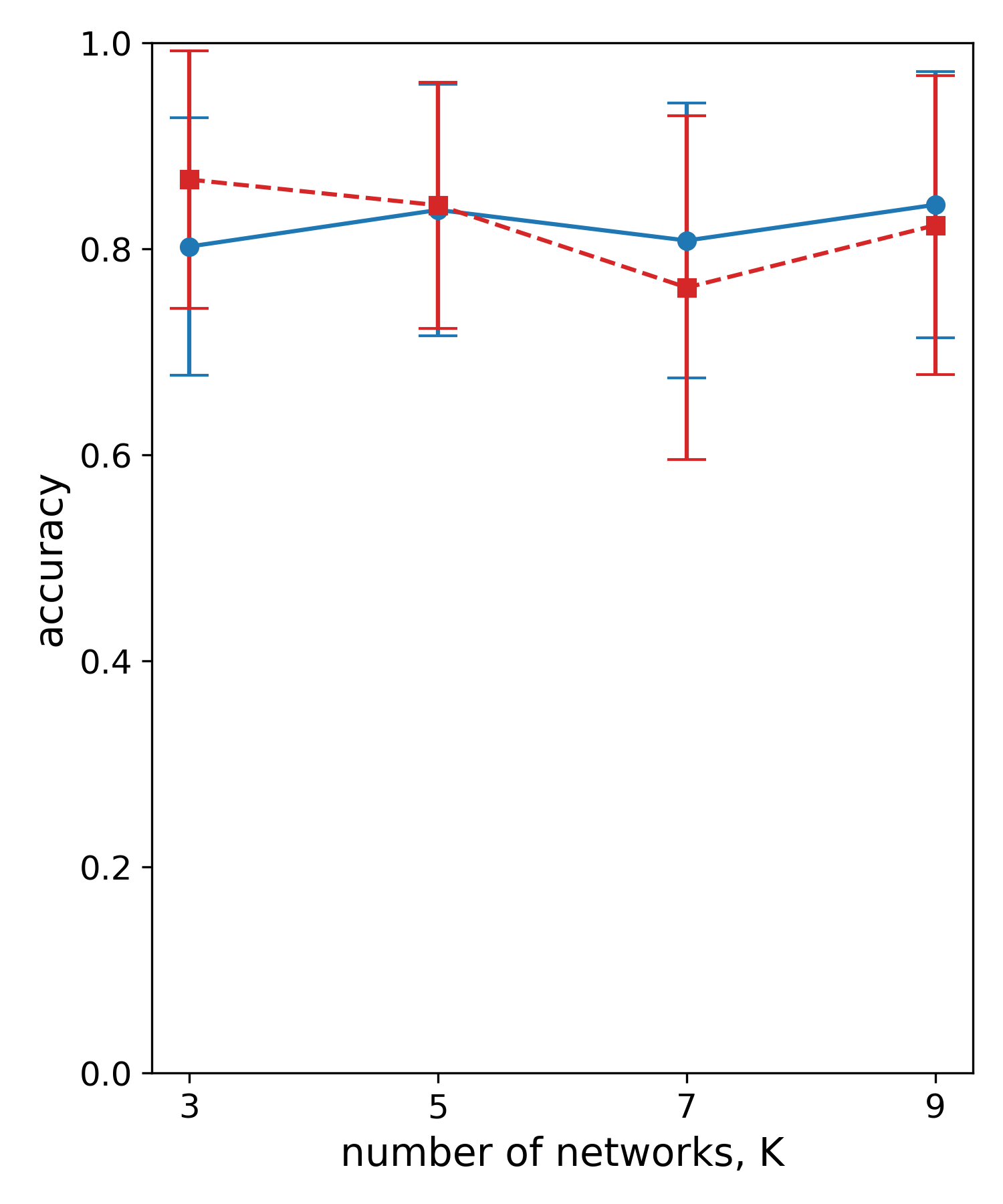}
 \caption{Accuracy of the CoOL-trained convolutional networks on a hold-out set of 5 MINST digits as a function of the number of networks.  Training is stopped when the agreement rate reaches 99.5\% or 5000 epochs, which ever comes first.}
\label{fig:K_ablation}
\end{figure}

The results of these parameter sweeps are shown in Figure \ref{fig:computation}.
We see that, as expected from Section \ref{sec:learning}, memory consumption increases linearly as a function of the number of samples during the batch training scheme.
The memory allocated is dominated by the gradients needed for backpropagation in the neural networks themselves, not in the EM portion, which is a tiny fraction of the total parameters.
The training time is also linear in the number of samples.
Using a mini-batch version places an upper boundary on memory consumption and also reduces training time.
The latter is a common benefit to mini-batch learning \cite{goodfellow_deep_2016}.
In Figure \ref{fig:K_ablation} we see that the number of networks, K, does not effect the performance of CoOL on hold-out data.

Training IDEC on the datasets used here takes a trivial amount of time.
However, the network used was pretrained extensively on the entire MNIST dataset wherein ``[e]ach layer [of the autoencoder] is pretrained for 50000 iterations with a dropout rate of 20\%. The entire deep autoencoder is further finetuned for 100000 iterations without dropout'' \cite{xie_unsupervised_2016}.
Using the same hardware as the CoOL cases and pre-loading the data on to the GPU, we find that 100 epochs of pretraining takes 96.9$\pm$1.8~seconds.
Thus, the 100000 epochs of finetuning alone would take 1615 minutes, far longer than any of the runs shown in Figure \ref{fig:computation}.

The training time and memory consumption for TELL are similar to CoOL, as is its performance.
On the other hand, SpectralNet trains faster and uses less memory because it approximates the affinity matrix in order to enable ``its scalability and thus allows one to cluster large datasets that are prohibitive for standard spectral clustering'' \cite{shaham_spectralnet_2018}.
However, SpectralNet was the poorest performer in terms of accuracy (see Table \ref{tab:mnist_success}).


\subsection{Ablation Study}
\label{app:ablation}

Herein we investigate the learning properties of CoOL by making some critical changes to the algorithm.
We vary the values of $\lambda$ and $\alpha$, defined in Section \ref{sec:learning}.
We substitute KL-divergence for the cross-entropy term in Eq. \ref{eq:full_loss}.
And, finally, we use a pretrained backbone to replace the randomly initialized networks.

\begin{figure}
 \centering
        \includegraphics[width=0.5\textwidth]{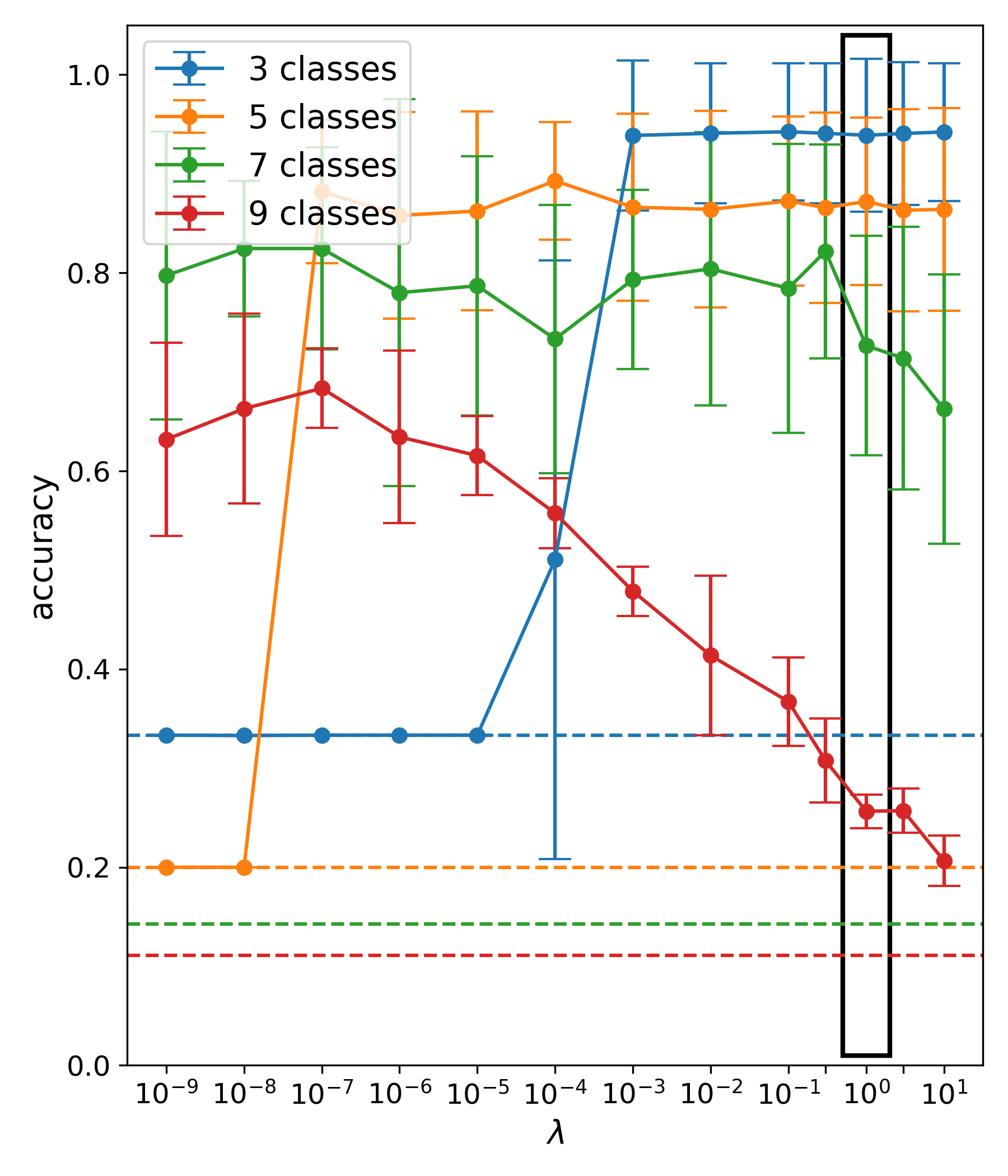}
 \caption{Accuracy of the CoOL-trained convolutional networks on a hold-out set of MINST images as a function of the weight of the determinant term, $\lambda$.  The box shows the settings used to produce Table \ref{tab:mnist_success}.  The dashed lines show the accuracy of a random guess model, where the colors match the colors shown in the legend.}
\label{fig:lambda_ablation}
\end{figure}

The first ablation we make is over $\lambda$.
We train on the MNIST dataset in the same fashion of section \ref{sec:mnist}, fix $\alpha=1$, and train until agreement reaches 99.5\% or 5000 epochs, which ever comes first.
We perform each experiment 3 times.
A plot of hold-out set accuracy versus $\lambda$ is shown in Figure \ref{fig:lambda_ablation}.
The performance depends on the selection of $\lambda$, but the peaks are broad so that very accurate selection of lambda is unnecessary.
In other words, a very coarse scan, perhaps a decade or two per step, appears sufficient to identify the region of best performance.
As usual, we monitor both agreement rate and the number of unique classes predicted.
It is easy to tell the difference between successful and unsuccessful runs.
Successful runs have smoothly increasing agreement rate that begins improving within a few hundred epochs of start and a stable number of unique classes predicted and unstable runs are missing one or both of these features.


Using the $\lambda$ values associated with the highest agreement rate, we vary $\alpha$ over many decades, including $\alpha=0$.
We find that $\alpha$ has no impact at all on the 3- and 5-class cases, and a modest, $\sim10\%$, effect on the higher class cases.

\begin{figure}
 \centering
        \includegraphics[width=0.5\textwidth]{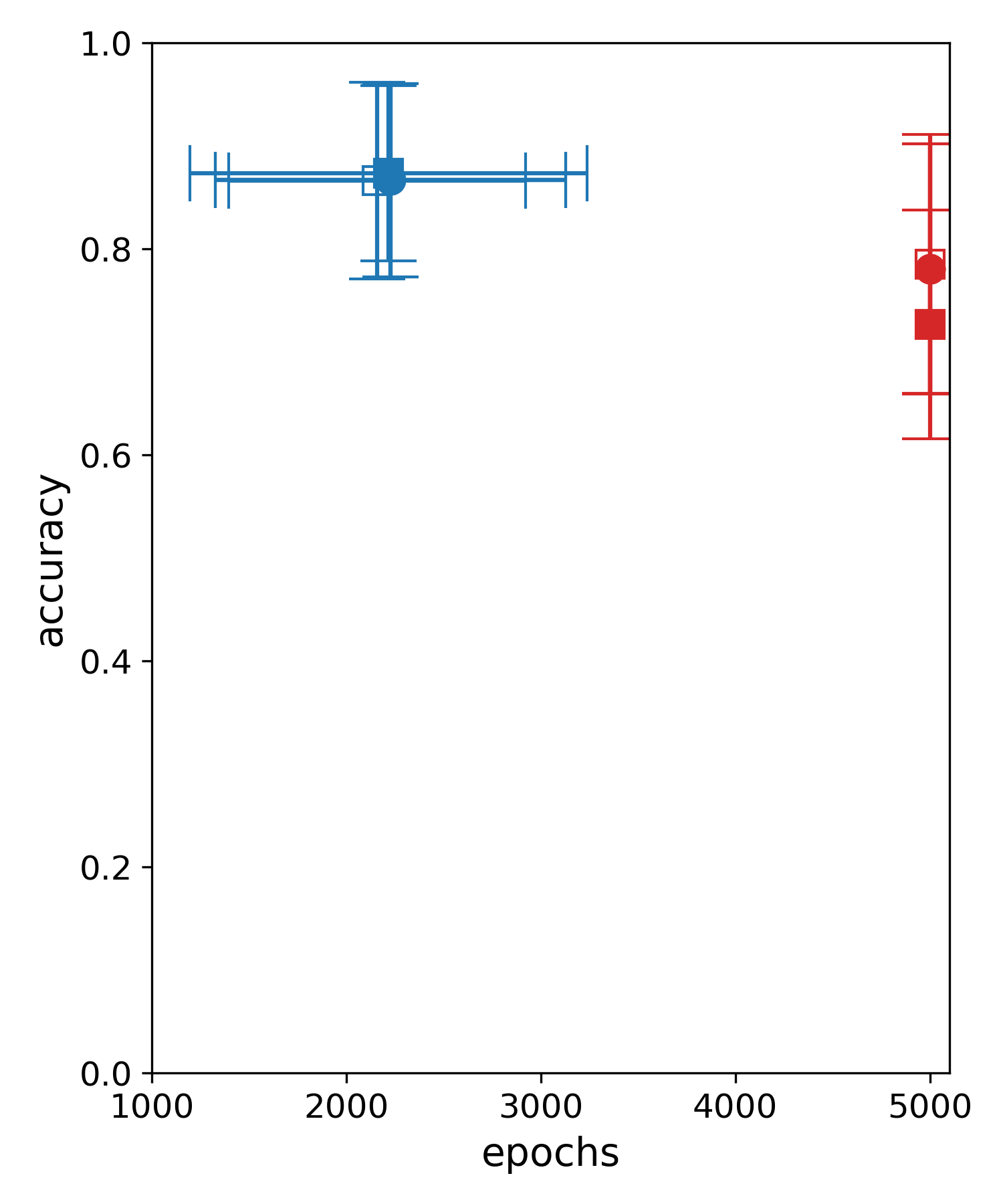}
 \caption{Accuracy of the CoOL-trained convolutional networks on a hold-out set of MINST images as a function of the number of epochs.  Training is stopped when the agreement rate reaches 99.5\% or 5000 epochs, which ever comes first.  The empty symbols represent models trained using Kullback–Leibler divergence without sampling and the full symbols are models trained using negative log-likelihood loss as in Section \ref{sec:learning}.  The blue symbols are for learning 5 digits and the red symbols are for 7 digits.  The circles represent $\alpha=0$ and the squares represent $\alpha=1$.}
\label{fig:kld}
\end{figure}

In Figure \ref{fig:kld} we compare the training time, measured in epochs, and the accuracy if we replace the negative log-likelihood loss used in Section \ref{sec:learning} with the Kullback–Leibler divergence.
We treat the model outputs as the approximating distribution and $T$ (see Equations \ref{eq:firstT} and \ref{eq:t_ij_update}) as the ``true'' distribution.
We also disable this term in the loss function by taking $\alpha$ as 0 (disable) or 1 (enable) and do the same for the negative log-likelihood loss.
Training is stopped when the agreement rate reaches 99.5\% or 5000 epochs, which ever comes first and we run all experiments 3 times.
This figure shows that the loss term does not change the learning as all of the results (for a particular number of classes) are clustered much more tightly than the error bars.

\begin{table}
\caption{
Performance of CoOL on the CIFAR10 dataset. 
The models are selected by highest agreement rate among all models that produce the correct number of unique labels.  
The columns show the mean and standard deviation of 3 runs for each dataset for all the indicated quantities.
The CoOL model converged only once on the 7-CIFAR dataset (thus, the lack of standard deviation) and never on the 9-CIFAR dataset.}
\centering
\begin{tabular}{| l l || c c c c |}
\hline
& & 3-CIFAR10 & 5-CIFAR10 & 7-CIFAR10 & 9-CIFAR10 \\
\hline
CoOL & agr. rate & 0.999$\pm$0.001 & 0.998$\pm$0.001 & 0.207 & \\
CoOL & acc. & 0.497$\pm$0.021 & 0.311$\pm$0.013 & 0.221 & \\
CoOL & NMI & 0.079$\pm$0.019 & 0.055$\pm$0.002 & 0.039 & \\
CoOL & ARI & 0.079$\pm$0.017 & 0.043$\pm$0.001 & 0.024 & \\
\hline
CoOL-ResNet50 & agr. rate & 1.000$\pm$0.000 & 1.000$\pm$0.000 & 1.000$\pm$0.000 & 0.893$\pm$0.005 \\
CoOL-ResNet50 & acc. & 0.468$\pm$0.072 & 0.329$\pm$0.007 & 0.263$\pm$0.031 & 0.235$\pm$0.008 \\
CoOL-ResNet50 & NMI & 0.078$\pm$0.029 & 0.068$\pm$0.007 & 0.088$\pm$0.008 & 0.093$\pm$0.006 \\
CoOL-ResNet50 & ARI & 0.082$\pm$0.034 & 0.053$\pm$0.006 & 0.061$\pm$0.009 & 0.055$\pm$0.009 \\
\hline
\end{tabular}
\label{tab:cifar_success}
\end{table}

To investigate the effect of a pre-trained backbone on CoOL's performance, we switch to the CIFAR-10 dataset \cite{krizhevsky2009learning}.
This is a much more difficult clustering problem.
We select the classes and samples in the same way as the MNIST experiments.
To replicate what a user might do with a dataset with unknown labels, we design a network that appears to learn via increasing agreement rate (available in the code that accompanies this paper).
We also mean-subtract and scale the input channel data such that all the input values lie in [-1, 1].
For the pre-trained backbone we use the ResNet-50 network available in the torchvision package \cite{he2015deepresidual, torchvision2016} and train only the final connections to the clustering output.
The input data is scaled in the same way that the data was during pre-training.
For both models, we train in the same way as we did for the MNIST experiments while scanning $\lambda$ from $10^{-9}$ to $10^6$ in steps of 10.
Model selection is done by first discarding any models that do not achieve the desired number of classes on output, and then selecting the model with the highest agreement rate.
This procedure can be used on any dataset without labels.

Performance of the selected models is shown in Table \ref{tab:cifar_success}.
We see that the arbitrary network that we created is partially successful at separating the different classes (in terms of accuracy).
However, it does not do well on the experiments with larger number of classes in that it seldom converges in the allowed number of epochs.
This serves as an indication that the models used are likely under-powered for the clustering task at hand.
The pre-trained feature-extractor and clustering head is much more successful as it converges on all data sets and out-performs a random guess model.
As pre-trained models are rare outside of well studied data types, such as the images we use here, we merely comment that pre-trained models are a good starting place, if one is available.

Summarizing the results of this appendix, including some details we uncovered in Section \ref{sec:description}, the following is a strategy for getting the best performance out of CoOL on new datasets without labels: 
(1) fix learning rate to $10^{-4}$ and do not use weight decay; 
(2) scan $\lambda$ over many decades and choose the value that maximizes agreement rate; 
(3) scan $\alpha$ over many decades and look for the value that maximizes agreement rate further still (It is also fine to take $\alpha=0$.);
(4) train on a large number of classes (say, 7 to 9) and check to see that the determinants of their outputs are smoothly increasing, if they are not, the networks might be under-powered for the clustering task.

\subsection{Groups of MNIST digits}
\label{app:mnist_groups}

Herein, we cluster all 10 digits of the MNIST handwritten digit dataset using only 4 clusters.
We train for 2000 epochs using 5 observers, 1000 samples per digit, a learning rate of $10^{-4}$, and the Adam optimizer.
This training is run 5 times to generate the groups shown in Table \ref{tab:groups_10}.
At first look, it appears that CoOL has not done anything sensible as, unlike Table \ref{tab:groups_1}, the groups all contain multiple digits and have relatively poor consistency.
However, the task of clustering the 10 digits into 4 clusters cannot possibly partition the samples into the digits.
Let us see what CoOL has found, instead.

\begin{table}
\caption{Table of 10 largest groups found when clustering 10 digits into 4 classes using CoOL.  The grouping produced 276 groups total.  Group is the group name for a set of samples; count is the number of samples in that group; label is the most common label in the group; labels is a list of all the labels found in the group; and consistency is the fraction of the group samples that have the label given in the label column.  A consistency of 1 means that all of the group samples have the same label.}
\centering
\begin{tabular}{l r c c c}
\hline
Group & count & label & labels & consistency \\
\hline
(2, 2, 0, 1, 0) & 1228 	& 1 & (0 1 2 3 4 5 6 7 8 9)  & 0.683 \\
(3, 3, 3, 0, 3) & 856 	& 0 & (0 2 3 4 5 6 7 8 9) 	 & 0.849 \\
(1, 0, 2, 3, 1) & 479 	& 7 & (1 3 4 7 8 9)      	 & 0.697 \\
(0, 0, 2, 3, 1) & 424 	& 4 & (3 4 7 8 9)        	 & 0.491 \\
(0, 1, 1, 0, 2) & 354 	& 6 & (0 2 3 5 6 8)      	 & 0.949 \\
(1, 0, 2, 2, 1) & 336 	& 9 & (2 4 7 9)          	 & 0.426 \\
(3, 0, 3, 3, 2) & 226 	& 2 & (0 2 3 4 9)        	 & 0.969 \\
(1, 1, 1, 2, 3) & 216 	& 3 & (1 3 4 5 8 9)      	 & 0.542 \\
(1, 1, 1, 0, 3) & 195 	& 3 & (2 3 4 5 8 9)      	 & 0.800 \\
(3, 3, 3, 0, 2) & 181 	& 6 & (0 2 3 4 6)        	 & 0.420 \\
(1, 3, 3, 0, 3) & 165 	& 4 & (3 4 5 7 8 9)      	 & 0.339 \\
\hline
\end{tabular}
\label{tab:groups_10}
\end{table}

\begin{figure}
 \centering
        \includegraphics[width=1.00\textwidth]{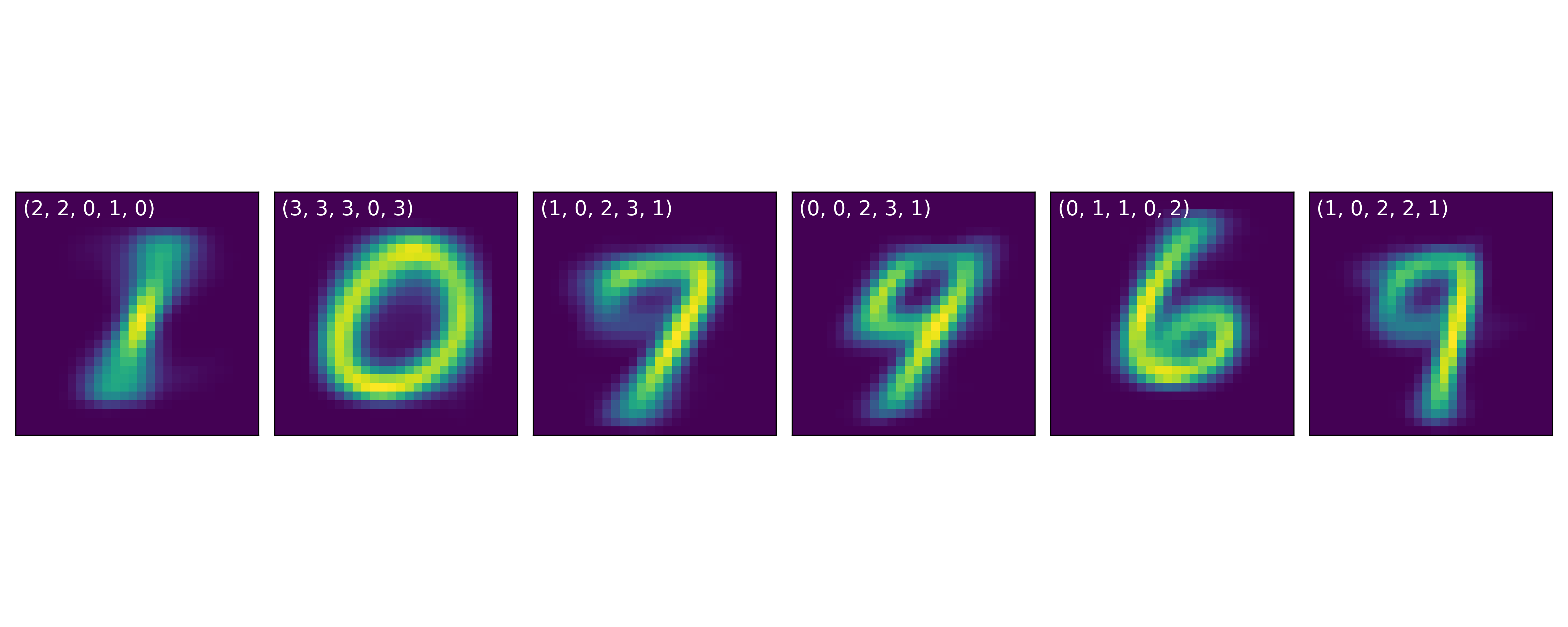}
 \caption{Composite images for the 6 largest groups shown in Table \ref{tab:groups_10}.}
\label{fig:composite_digits}
\end{figure}

By summing all the images on a per-group basis, we can plot composite images of the ``mean'' image for each group.
This is shown in Figure \ref{fig:composite_digits}.
We see that the first six groups contain images of digits that look like 1, 0, 7, 4, 6, and 9 (from left to right in the figure).
We also see that the three groups that are near each other make digits that look very similar to each other.
Group (1, 0, 2, 3, 1) looks more like a 7, group (0, 0, 2, 3, 1) looks more like a 4, and group (1, 0, 2, 2, 1) looks more like a 9, but all three composite images look very similar is well.

We include this appendix to demonstrate that guessing the correct number of semantic labels in the data set is not necessary to get useful information (in this case digit-like representations as shown in Figure \ref{fig:composite_digits}) from CoOL, or, indeed, any clustering algorithm.

\ack{The authors wish to acknowledge useful conversations with Eric Darve and Daniel Ratner.}

\funding{
SLAC National Accelerator Laboratory is supported by the U.S. Department of Energy, Office of Science under Contract No. DE-AC02-76SF00515.
This research used resources of the National Energy Research Scientific Computing Center, a DOE Office of Science User Facility supported by the Office of Science of the U.S. Department of Energy under Contract No. DE-AC02-05CH11231.
}

\roles{FHO was responsible for the conceptualization, formal analysis, investigation, methodology, software, validation, visualization, and writing. MEM was responsible for funding acquisition, project administration, resources and supervision.}

\data{
The MNIST dataset is available here \cite{lecun_mnist_1998} and the CIFAR-10 dataset is available here \cite{krizhevsky_cifar_2009}.
The authors retrieved the data through torchvision \cite{torchvision2016}.
}


\suppdata{
} 


\bibliographystyle{ieeetr}
\bibliography{main}

\end{document}